\documentclass[reprint,amsmath,amssymb,aps,floatfix,superscriptaddress
]{revtex4-1}

\usepackage{graphicx}
\usepackage{dcolumn,xcolor,hyperref}
\usepackage{bm}

\begin{document}


\title{Physics-tailored machine learning reveals unexpected physics in dusty plasmas}

\author{Wentao Yu}
\affiliation{Department of Physics, Emory University, 400 Dowman Dr., Atlanta, GA 30322, USA}
\author{Eslam Abdelaleem}
\affiliation{Department of Physics, Emory University, 400 Dowman Dr., Atlanta, GA 30322, USA}
\author{Ilya Nemenman}
\affiliation{Department of Physics, Emory University, 400 Dowman Dr., Atlanta, GA 30322, USA}
\affiliation{Department of Biology and Initiative for Theory and Modeling of Living Systems, Emory University, 1510 East Clifton Road NE, Atlanta, GA 30322, USA}
\author{Justin C. Burton}
\email{justin.c.burton@emory.edu}
\affiliation{Department of Physics, Emory University, 400 Dowman Dr., Atlanta, GA 30322, USA}

\date{\today}

\begin{abstract}
Dusty plasma is a mixture of ions, electrons, and macroscopic charged particles that is commonly found in space and planetary environments \cite{merlino2021dusty}. The particles interact through Coulomb forces mediated by the surrounding plasma, and as a result, the effective forces between particles can be {\color{black} non-conservative and non-reciprocal}. Machine learning (ML) models are a promising route to learn these complex forces, yet their structure should match the underlying physical constraints to provide useful insight \cite{carleo2019machine}. Here we demonstrate and experimentally validate an ML approach that incorporates physical intuition to infer force laws in a laboratory dusty plasma. Trained on 3D particle trajectories, the model accounts for inherent symmetries, non-identical particles, and learns the effective non-reciprocal forces between particles with exquisite accuracy (\textit{R}$^2>$ 0.99). We validate the model by inferring particle masses in two independent yet consistent ways. The model's accuracy enables precise measurements of particle charge and screening length, discovering large deviations from common theoretical assumptions. Our ability to identify new physics from experimental data demonstrates how ML-powered approaches can guide new routes of scientific discovery in many-body systems. Furthermore, we anticipate our ML approach to be a starting point for inferring laws from dynamics in a wide range of many-body systems, from colloids to living organisms \cite{pineda2023geometric,bruckner2021learning,toner1998flocks}.  
\end{abstract}

\maketitle

\section{Introduction}
Dusty plasma is ubiquitous throughout the universe, from Saturn's rings to interstellar space \cite{melzer2019physics,hippler2008low,chaudhuri2011complex,merlino2021dusty}, and is critically important for planet formation \cite{goertz1989dusty, wahlund2009detection,shukla2002dust}, technological processes \cite{kortshagen2009nonthermal, merlino2004dusty,beckers2019euv,winter2000dust}, and potentially the emergence of life \cite{tsytovich2007plasma}. In a dusty plasma, particle interactions have known approximations based on tractable physics, yet they are poorly understood in environments that deviate from the simplest equilibrium conditions, for example, in systems with background plasma flows \cite{matthews2020dust} or with external magnetic fields \cite{melzer2021physics,thomas2012magnetized}. Particles interact through 
complicated forces mediated by the plasma environment \cite{melzer2008fundamentals}, and violate some of our basic expectations: they are non-reciprocal {\color{black}and can source energy from their nonequilibrium environment} \cite{melzer2019finite,nikolaev2021nonhomogeneity,kolotinskii2021effect,vaulina2015energy,ivlev2015statistical}. Limited information about these interactions can be obtained by carefully investigating quiescent systems of particles, for example, the Brownian motion of two particles \cite{ding2019nonlinear,mukhopadhyay2012two,ding2021machine,yu2022extracting} or the vibrational modes in a strongly-coupled crystal \cite{nunomura2002dispersion,couedel2009first,couedel2010direct,zhdanov2009mode}. Yet particles must be highly dynamic and explore phase space to learn a separation-dependent interaction law  \cite{konopka2000measurement,gogia2017emergent}. 
Thus, compact and precise mathematical expressions that summarize interactions among dust particles as physical laws do not exist, yet some constraints on the interactions are clear. For example, the forces between particles are expected to be pairwise {\em to leading order} and to depend only on their mass, charge, and the spatial configuration \cite{greiner2018diagnostics,lampe2000interactions,gopalakrishnan2012coulomb,ignatov1997interaction}. To handle this complexity, here we introduce a broadly-applicable ML approach to infer new, previously unknown interactions in dusty plasmas. Our approach incorporates physical constraints in its underlying neural network architecture to learn the external forces and the unknown particle interactions directly from experimental data.

Broadly speaking, dusty plasma is a many-body system of interacting particles. Many-body systems are abundant in nature and continue to push the boundaries of science, from the detection of exoplanets \cite{fulton2018radvel,mayor1995jupiter} to the behavior of living organisms \cite{pineda2023geometric,bruckner2021learning,toner1998flocks}. In these systems, interaction laws are often not well-defined, unlike Newton's laws of classical physics. However, the ability to generate large, precise data sets and the simultaneous emergence of machine learning (ML) to analyze them offer a path for inferring these interactions from experimental data. Many ML algorithms can model these complex systems by inferring parameters in a pre-defined mathematical description that best fit the data {\color{black}\cite{bapst2020unveiling, colen2021machine, tah2022fragility,daniels2019automated,anstine2023machine}}, or by finding a functional form describing the system within a constrained (though often large) library of possibilities {\color{black}\cite{champion2019data, brunton2016discovering, rudy2017data, bruckner2020inferring,lemos2023rediscovering}}. Other ML algorithms focus directly on predicting the future state of a system from its past without inferring or interpreting the underlying physics as an intermediate step \cite{bruckner2021learning,pandarinath2018inferring,colen2021machine,daniels2015automated, chen2022automated}. Often the data used to train and validate these models come from simulations with labeled ground-truth parameters, known particle properties, and provided, well-defined interaction laws. However, real experimental data lacks all of these conveniences, and there have been recent attempts to extend ML methods to experimental data \cite{supekar2023learning,raissi2020hidden,chen2022automated,bruckner2021learning,colen2021machine,ruiz2024discovering}. Nevertheless, endowing ML methods with an inductive bias based on physical intuition can facilitate progress in realistic situations. This is especially important for many-body data, where such constraints are needed to tame the combinatorial complexity of interactions among the measured components \cite{battiston2021physics}, and as a result, physics-constrained machine learning for many-body systems is still emerging \cite{carleo2019machine, karniadakis2021physics,bapst2020unveiling,cichos2020machine,falk2021learning,han2022learning,ruiz2024discovering}. Here we simultaneously address many of these challenges by introducing a physics-constrained ML approach based on neural networks as universal approximators, which is able to learn {\em new}, unanticipated interaction laws dusty plasma experiments.

\begin{figure*}[t]
\centering
\includegraphics[width=.9\textwidth]{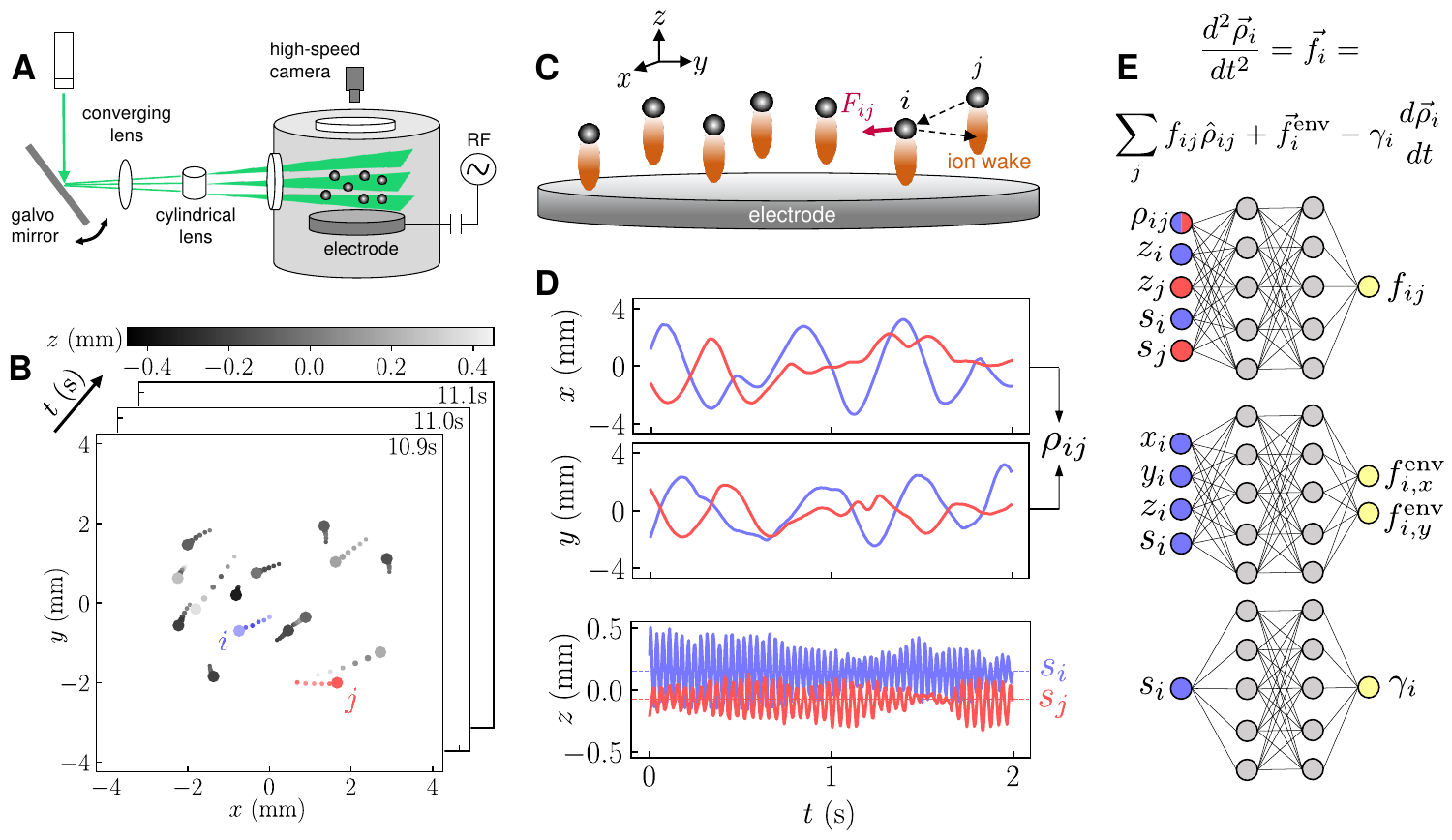}
\caption{{\color{black}Overview of experiment and data workflow. ({\bf A}) Charged microparticles are levitated in an RF-driven plasma sheath above a flat electrode. Their motion is imaged using a scanning laser sheet coupled to a high-speed camera \cite{yu20233d}. ({\bf B}) Snapshot of particle positions from a single experiment of 15 particles. The grayscale color indicates the $z$-position, and the tails of each particle represent the previous 5 frames. ({\bf C}) The focused ion wake (red) is directly below each particle, and contributes a small attractive part of the total force ($F_{ij}$) on particle $i$, so that the overall interaction is nonreciprocal. ({\bf D}) The $x$, $y$, and $z$ position of two particles during two seconds. The particles are marked $i$ (blue) and $j$ (red) in panel (C). The quantity $s_i=\langle z_i\rangle$ is used as a size identifier for each particle. ({\bf E}) The objective is to infer the horizontal reduced forces on particles using Newton's equation of motion. The schematic of the model, which consists of three neural networks trained concurrently and act as nonlinear approximators to the three terms in the equation (particle interactions -- $f_{ij}$, environmental forces -- $\vec{f}_\text{env}$, and damping from the background gas -- $\gamma_i$). The input color designates the source (particle $i$ or $j$).}}
\label{p1}
\end{figure*}

In a dusty plasmas,
the equilibrium charge, $q$, on a given particle is determined by a balance of ion and electron currents to the surface, and the currents are determined by the local plasma environment (ion and electron densities and velocity distributions). Usually particles acquire a negative charge, and thus repel. However, the interaction force is screened by the plasma environment, and the screening length $\lambda$ is of fundamental importance since it determines the effective range of interaction. Moreover, when particles are levitated in a plasma sheath near a conducting wall (electrode, Fig.~\hyperref[p1]{1\textit{A-C}}), $q$ will be a function of the particle's vertical $z$ position within the sheath. Particles also experience a fast-flowing ``river'' of ions that produces ion wakes behind each particle \cite{ishihara1997wake,ludwig2012wake}, and the effective interactions between particles include this ion wake (Fig.~\hyperref[p1]{1\textit{C}}). {\color{black}This wake-mediated interaction is non-reciprocal, breaks translational symmetry in $z$, and is predicted to cause attractive forces when particles are close.} Since dusty plasmas are readily confined and manipulated in the laboratory, they offer an ideal platform to study complex and emergent collective behavior in particulate matter. 

To infer interaction laws in dusty plasma, we captured three dimensional (3D) trajectories of individual dust particles using scanning laser sheet tomography \cite{yu20233d}. Our physics-constrained neural network model used these trajectories to infer non-reciprocal interactions between individual pairs of non-identical particles, environmental forces that trap particles and drive their motion, and velocity-dependent drag forces from the background gas. The inference procedure compared the sum of these forces to the experimental acceleration of each particle. Remarkably, the model was extraordinarily accurate when fitting the acceleration, achieving $R^2> 0.99$ over multiple experiments. 
To verify that the model learned each force accurately, we compared the mass, $m$, of each particle in two independent ways, which agreed with each other (and the known size of the particles using optical microscopy). For particles in the same horizontal plane, we fitted the interaction force of each particle pair to a well-known analytical approximation \cite{konopka2000measurement,melzer2008fundamentals}, allowing us to simultaneously extract $m$, $q$, and $\lambda$. Contrary to conventional assumptions where $\lambda$ depends solely on plasma properties, we find that the fitted value of $\lambda$ increases with the average size of interacting particles. Furthermore, we find that $q\sim m^{p}$, where $p$ ranges between 0.30 and 0.80 and increases with background gas pressure. This variation contrasts with the simplest assumptions of particle charging in dusty plasmas where $q\propto m^{1/3}$ \cite{melzer2019physics,goree1994charging}. 

\section*{Experiments and model} 

{\color{black}Our dusty plasma experiments utilized a vacuum chamber filled with argon gas at 0.5-1.5 Pa of pressure (Fig.~\hyperref[p1]{1\textit{A}}) -- a setup similar to our previous experiments \cite{gogia2017emergent,yu2022extracting,yu20233d,harper2020origin}. A disk-shaped electrode was driven with RF power and generated a weakly-ionized argon plasma. Near the electrode surface, micron-sized charged particles were levitated in a plasma sheath -- a sharp gradient in the electric field where electrostatic forces can balance with gravitational forces \cite{nitter1996levitation}. The levitated particles explored a space roughly 10 mm $\times$ 10 mm $\times$ 1 mm in size (Fig.~\hyperref[p1]{1\textit{B}}, Movie S1). The particle positions were tracked using 3D scanning laser tomography \cite{yu20233d}. For more details about the plasma conditions and particle tracking method, see \textit{Materials and Methods}. 

The interaction between each pair of particles is mostly due to electrostatic repulsion since each particle carries a negative charge ($\approx10^4$e). However, in the plasma sheath, ions stream past each particle at speeds greater than 2 km/s, resulting in vertical ion wakes (Fig.~\hyperref[p1]{1\textit{C}}), similar to the wakes produced behind a fast-moving boat in open water. The wakes make the overall interactions between particles nonreciprocal, i.~e., $F_{ij}\neq F_{ji}$ \cite{melzer2019finite,nikolaev2021nonhomogeneity,kolotinskii2021effect,vaulina2015energy,ivlev2015statistical}, and they are especially important when particles are vertically separated in the $z$ direction. Moreover, since the particle charge varies within plasma sheath, the particle interactions break translational symmetry in $z$, while maintaining translational symmetry in the $xy$-plane. To learn these complex forces, we overcome a major challenge: building the required physical symmetries into a model that can be trained on systems with varying particle number.   
}
 
The tracked 3D trajectories ($x_i(t)$, $y_i(t)$, $z_i(t)$) of all the particles were used as input to train our ML model. An example of trajectories for two particles is shown in Fig.~\hyperref[p1]{1\textit{D}}. The model assumes that the horizontal ($xy$-plane) acceleration of each particle is determined by Newton's 2nd law:
\begin{equation}
    \Ddot{\vec{\rho}}_i = \vec{f}_i = \sum_{j\neq i} f_{ij} \hat \rho_{ij} + \vec{f}^{\text{env}}_i - \gamma_i \dot{\vec{\rho}}_i,
    \label{model}
\end{equation}
where $\vec f_i$ is the horizontal reduced force on particle $i$, or equivalently the net force, $\vec F_i=(F_{i,x},F_{i,y})$, divided by its mass, $m_i$. Dotted variables represent differentiation with respect to time. The position and displacement vectors are $\vec\rho_i = (x_i, y_i)$ and $\vec\rho_{ij} = (x_i - x_j, y_i - y_j) = \rho_{ij}\hat{\rho}_{ij}$, where $\hat{\rho}_{ij}$ is the direction of the reduced horizontal interaction force from particle $j$ to $i$, and $f_{ij} = F_{ij} / m_i$, where $F_{ij}$ is the magnitude of the force. Since the ion wake is directly below each particle, as shown in Fig.~\hyperref[p1]{1\textit{C}}, the ion wake will change the direction of the $z$-component of the force, but interaction forces in the $xy$-plane will still point along $\hat{\rho}_{ij}$ \cite{ivlev2015statistical}. The reduced environmental force is $\vec{f}^{\text{env}}_i = \vec{F}^{\text{env}}_i/m_i$, where $\vec{F}^{\text{env}}_i$ is the horizontal environmental force on particle $i$, and the damping coefficient of particle $i$ is $\gamma_i$. Particles are strongly confined by gravity and electrostatic forces in the $z$-direction, which are about 100 fold larger than other forces in the system, as evidenced by the different frequencies and amplitudes of motion shown in Fig.~\hyperref[p1]{1\textit{D}}. 

{\color{black} Although we can accurately track the $z$ position of each particle since the vertical oscillation frequency was $\approx$ 25 Hz (Fig.~\hyperref[p1]{1\textit{D}}) and our sampling rate was 200 Hz, inferring forces in our model requires integrating the data over a small time window (see \textbf{SI}, Eq.~S5), which would necessitate a higher time resolution for $z$ force inference. Thus, in this study, we only aim to infer forces in the $xy$-plane, which will generally depend on the $z$ position of each particle.} Importantly, the particles in our experiments were not identical, and the model requires particle-level identifiers. Ideally, this would be the mass of each particle, which is unknown. But heavier particles sit lower in the plasma sheath, and we found that a good  identifier ($s_i$) for the size of each particle was simply its mean $z$-position, averaged over an entire time series: $s_i = \langle z_i\rangle_t$. 

\begin{figure}[t]
\centering
\includegraphics[width=.9\columnwidth]{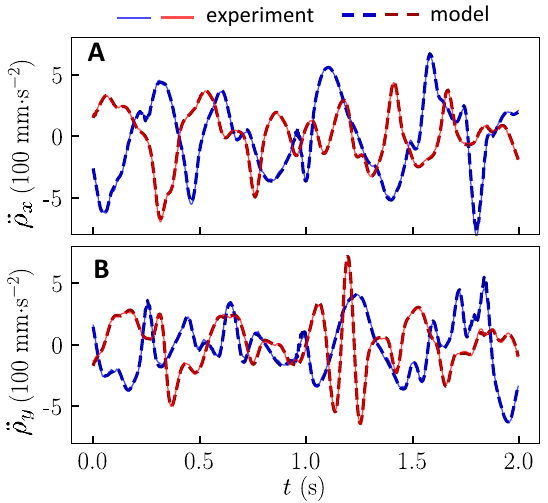}
\caption{The predicted reduced force ($\vec f$, dashed lines) and measured experimental acceleration ($\Ddot {\vec \rho}$, solid lines) for 2 particles (red and blue) in the 15 particle system. We note that this is \emph{test data}, meaning it was not used to train the model. Data is shown for 2 s out of the 4.94 s of test data. The entire experiment was 49.4 s long. ({\bf A}) $f_x$ and $\Ddot{\rho}_x$, and ({\bf B}) $f_y$ and $\Ddot{\rho}_y$. The two particles are the same particles shown in Fig.~\hyperref[p1]{1\textit{D}}.
\label{p2}}
\end{figure}

In the model, three neural networks (NNs) act as universal approximators to the three types of forces on each particle (Fig.~\hyperref[p1]{1\textit{E}}). {\color{black} We use three independent networks because they represent different terms in the equation of motion of a single particle, Eq.~\ref{model}. Each network must necessarily have different inputs, otherwise we would not be able to distinguish the learned forces from the three terms.} The first NN, $g_\text{int}$, requires $\rho_{ij}$, $z_i$, $z_j$, $s_i$, and $s_j$ as inputs. It outputs the magnitude of the effective reduced interaction force, $f_{ij}$. We note that this structure conserves translational symmetry in $x$ and $y$, but breaks this symmetry in $z$. 
The second NN, $\vec{g}_\text{env}$, requires $x_i, y_i, z_i, s_i$ as inputs. It outputs both components of the vector, $\vec{f}^{\text{env}}_i$. The third NN, $g_\gamma$, uses $s_i$ as its sole input, and outputs the drag coefficient, $\gamma_i$. Requiring a drag force linear in velocity is supported by theory: according to Epstein's law \cite{epstein1924resistance}, for spherical MF particles with a density of 1510 kg$\cdot$m$^{-3}$ inside argon gas \cite{melzer2008fundamentals}, 
\begin{equation}
   \gamma_i = \frac{12.2P}{d_i} \mu\text{m}\cdot\text{Pa}^{-1}\cdot\text{s}^{-1}.
    \label{epstein}
\end{equation}
Here $P$ is the plasma pressure and $d_i$ is the diameter of particle $i$. Inferring an individual particle's damping coefficient provides direct information about its size (and mass), thus $g_\gamma$ constructs a map from the size identifier $s_i$ to the physical parameter $\gamma_i$ (or $m_i$). 

During training, the model adjusts the weights in each neural network concurrently to minimize a loss function that compares the predicted reduced force, $\vec f_i$, to the measured horizontal acceleration, $\Ddot{\vec{\rho}}_i$. Since we are calculating the forces between all pairs of particles, the total training time scales as $N_p^2$. {\color{black} To reduce noise, we use the weak form in our loss function \cite{gurevich2019robust}, a technique that calculated a filtered version of acceleration from  experimental data by integrating trajectories over a small time window instead of computing derivatives from the noisy position time series (i.~e., Eq.~S5). 
As a simple example, consider the function centered at $t=0$: $w(t)=(1-t^2)^2$. Since $w(\pm 1)=0$ and $\dot{w}(\pm 1)=0$, it is straightforward to show using integration by parts that:
\begin{equation}
    \int_{-1}^1 \ddot{x}wdt=\int_{-1}^1 x\ddot{w}dt.
\end{equation}
This replaces a noisy second derivative of an experimental time series with exact derivatives of an analytic function ($w$). We found that this dramatically increased the model performance ($R^2$) and was a necessary part of our methodology.} The complete details of the model structure, minimization of the loss function, and the application of the weak form are described in supporting information (\textbf{SI}).


\section*{Results and Discussion}
\label{rst}
{\color{black} The presentation of our results are organized as follows. We first demonstrate the model's accuracy in predicting experimental trajectories using the sum of the three force components in Eq.~\ref{model}. Next we show that model's prediction on position-dependent interaction and environmental forces (the first 2 terms on the right-hand side of Eq.~\ref{model}). We then subsequently fit the predicted interaction (reduced force) using a well-known theory to extract estimates of each particle's charge, mass, and pairwise screening length. Finally, we verify the mass estimate by directly comparing it with expectations from the damping coefficient (last term in Eq.~\ref{model}) and discuss deviations of the fitted charge and screening length from conventional plasma physics theory.}

\subsection*{Model accuracy in fitting acceleration}
\label{2a}
We used the model to infer forces on particles from 5 experiments (movies S1-S5) carried out under different conditions: number of particles, gas pressure, and plasma conditions. At least $\sim 9$ particles were necessary to produce a highly dynamic system; smaller systems with less particles tended to form rotating crystalline structures (Movies S6-S7). We performed 10-fold cross-validation on each experiment. {\color{black} That is, the data, ordered by time, was split into 10 equal parts. We created 10 independent models where the $k$-th model ($k$ = 1,2,\dots,10) used the $k$-th part as the validation set while the other 9 parts were used for training. The model's performance was evaluated by computing the $R^2$ score on the validation set. The errors presented in our results represent the standard deviation of the prediction from all 10 models. Further details can be found in the {\bf SI}.} 

For each experiment, the average $R^2$ of the 10 models was always larger than 0.99 (Table \ref{r2}). For visual reference of the model performance, we show data for the $x$ and $y$ acceleration on two different particles and the corresponding model prediction in Fig.~\hyperref[p2]{2\textit{A-B}}. This remarkable agreement is representative of all 49.4 s of data captured in the experiment. We note that a high $R^2$ only indicates that the model fits the {\em sum} of the three reduced force components in Eq.~\ref{model}, and does not necessarily indicate that {\em each} component is fit correctly. Thus, we ensured that the set of input parameters for each component was parsimonious and contained minimal overlap, i.e., $x_i$ and $y_i$ appear directly as inputs to $\vec{g}_{\text{env}}$, but only appear in the particle separation $\rho_{ij}$ for $g_{\text{int}}$. Furthermore, as we will show, the accuracy of each component is validated by inferring particle-level properties in two independent ways.

\subsection*{Model prediction for each force component}
\label{2b}
Recent examples using graphical neural networks show that effective \emph{local} interaction forces can be learned from experimental data by assuming all particles are identical, and computing the average force \cite{ruiz2024discovering}. Underdamped Langevin inference (ULI) can also extract complex interactions between identical particles \cite{bruckner2020inferring}. In contrast to these examples, our model predicts the effective reduced interaction force, $f_{ij}$, which can be non-reciprocal, between \emph{any} particle pair $i$ and $j$ at \emph{any} position represented in the experimental data.  We are not aware of any other force inference technique that is capable of treating particles as individuals.

\begin{figure}[t]
\centering
\includegraphics[width=\columnwidth]{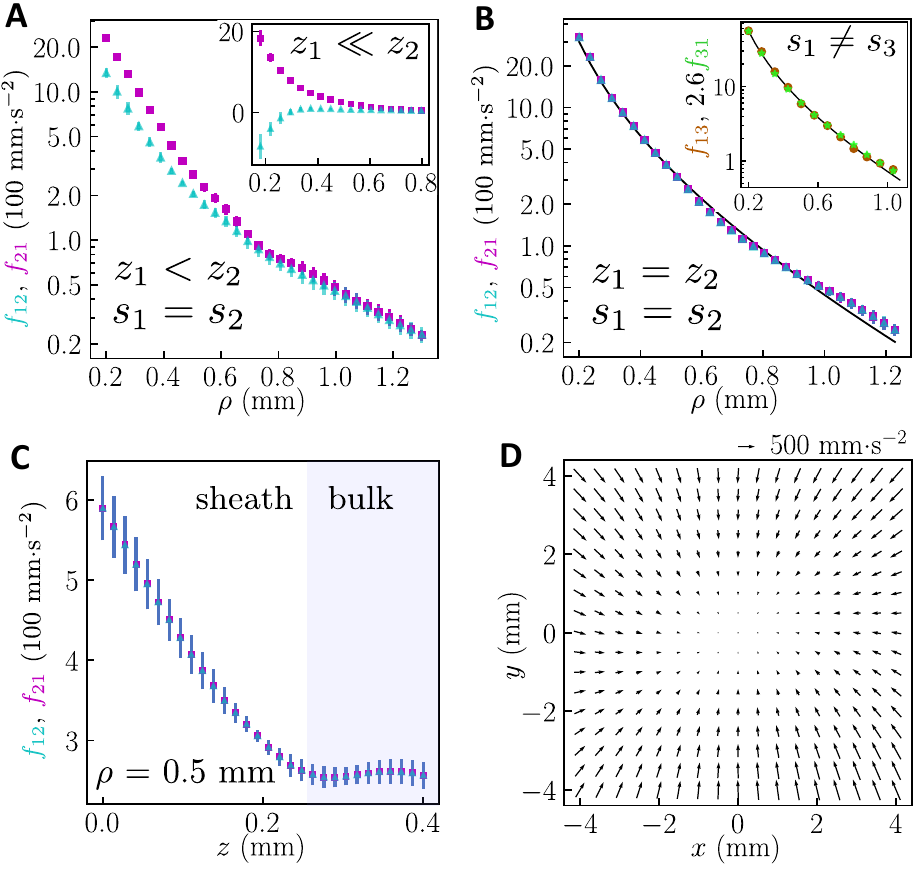}
\caption{Model prediction of interaction and environmental reduced forces for the 15-particle experiment. ({\bf A}) The magnitude of the reduced interaction force ($f_{12}$, cyan triangles; $f_{21}$, purple squares) between two similar particles ($s_1 = 0.234$ mm, $s_2 = 0.232$ mm), at $z_1 = 0.15$ mm and $z_2 = 0.30$ mm. The force is plotted versus the horizontal separation $\rho$. The inset shows the interaction at $z_1 = 0.05$ mm and $z_2 = 0.35$ mm. ({\bf B}) The model predicts the same two particles' interaction is reciprocal at $z_1 = z_2 = 0.15$ mm. The black solid line is a fit of the average of the two predictions to Eq.~\ref{yukawa} with $\lambda$ = 0.42 mm. The inset shows the interaction of two different particles ($f_{13}$, brown circles; $f_{31}$, green stars) at $z_1 = z_3 = 0.15$ mm. Here $s_3 = -0.053$ mm, and $f_{31}$ is shifted by a factor of 2.6 (the mass ratio) to collapse the curves. The black solid line is a fit to Eq.~\ref{yukawa} with $\lambda$ = 0.48 mm. ({\bf C}) $f_{12}$ and $f_{21}$ evaluated at $\rho = 0.5$, plotted versus $z = z_1 = z_2$. The sharp rise in the model prediction indicates the boundary between the plasma sheath and bulk plasma (purple). ({\bf D}) Environmental reduced force field of particle 1, $\vec{f}_1^{\text{env}}$, at $z_1 = 0.15$ mm. The error bars represent the standard deviation of the prediction from 10 models trained on different sections of the experimental data, as detailed in the \textbf{SI}.}
\label{p3}
\end{figure}

For simplicity, since $\rho_{ij} = \rho_{ji}$, we use $\rho$ to denote the horizontal separation of two particles. Figure~\hyperref[p3]{3\textit{A}} demonstrates the model's ability to capture non-reciprocal interactions for two nearly identical particles with identifiers $s_1 \approx s_2$ 
at different vertical positions, $z_1<z_2$.
Non-reciprocity is clearly observed for $\rho<$ 0.6 mm, and $f_{21}/f_{12}\approx 2$ at the shortest separation. For the same particles with a larger vertical separation, $f_{12}$ is attractive (Fig.~\hyperref[p3]{3\textit{A}} inset). The dramatic non-reciprocity is due to the presence of the ion wake structure beneath each particle, as shown in Fig.~\hyperref[p1]{1\textit{C}} \cite{vladimirov1995attraction}. However, interactions are expected to be reciprocal when $z_i=z_j$ \cite{ivlev2015statistical}. This reciprocity is illustrated in Fig.~\hyperref[p3]{3\textit{B}} for the same two particles (the main panel) and two different particles (inset). 

\begin{table}[htbp]
    \centering
        \caption{Parameters and model performance from 5 experiments. $N_p$ is the number of particles, $P$ is the neutral gas pressure, $z_\text{std}$ and $\rho_\text{std}$ are the standard deviation of the particle motion in the vertical and horizontal directions, respectively, and are averaged over all particles. Test $R^2$ is the $R^2$ score of the model performance on the test data set. Each experiment is assigned a color, indicated by the last column, which is plotted in Fig.~4.  }
    \begin{tabular}{@{}llllll@{}}
        \toprule
        $N_p$ & $P$ (Pa) & $z_\text{std}$\ (mm) & $\rho_\text{std}$\ (mm) & test $R^2$ & color \\ 
        9 & 1.00 & 0.060 & 0.96 & 0.9949&blue\\
        10 & 1.00 & 0.10 & 1.23 & 0.9921&green\\
        13 & 1.00 & 0.082 & 1.14 & 0.9912&red\\
        15 & 0.75 & 0.12& 2.24 & 0.9919&orange\\
        18 & 1.20 & 0.033 & 1.38 & 0.9963&purple\\
    \end{tabular}
    \label{r2}
\end{table}

In this reciprocal regime, we used the well-known screened Coulomb interaction to fit the prediction of the model:
\begin{equation}
    m_if_{ij} = m_jf_{ji} = \dfrac{A}{\rho}\left(\dfrac{1}{\rho} + \dfrac{1}{\lambda}\right)e^{-\rho/\lambda}.
    \label{yukawa}
\end{equation}
Here the coefficient $A$ is a fitting parameter, but theory suggests that $A = q_iq_j/4\pi\epsilon_0$, where $q_k$ and $m_k$ are the charge and mass of particle $k$, respectively, $\epsilon_0$ is the permittivity of free space, and $\lambda$ is the effective screening length \cite{melzer2008fundamentals,konopka2000measurement,melzer2019physics}. 
Importantly, systematic error can be clearly observed in the fit (solid lines in Fig.~\hyperref[p3]{3\textit{B}}), indicating that there are deviations from Eq.~\ref{yukawa} as a universal law for all particle separations. This deviation is expected since the real interaction involves both negatively-charged particles and their associated ion wake structures. These structures are often modeled as a virtual, positive charge below each particle \cite{kryuchkov2020strange}. Nevertheless, Eq.~\ref{yukawa} is a good analytical approximation for each pair of particles when they are at the same $z$, although as we will show, care must be taken when interpreting both $q$ and $\lambda$ from the fits to Eq.~\ref{yukawa}.  
When $z_i=z_j$, but $s_i\neq s_j$, as shown in the inset of Fig.~\hyperref[p3]{3\textit{B}} for different particles with indices 1 and 3, the reduced force can be shifted to coincide using a multiplicative factor of 2.6. This factor is the particles' mass ratio, $m_3/m_1$, when the forces are reciprocal ($F_{13}=F_{31}$). 

\begin{figure*}[t]
\centering
\includegraphics[width=.9\textwidth]{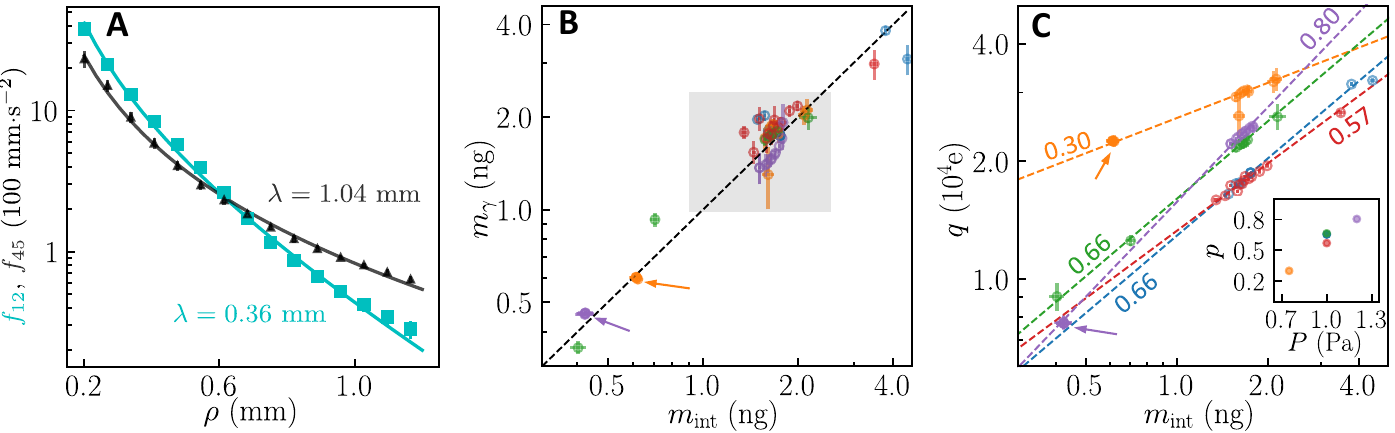}
\caption{The inferred measurements of mass, charge, and screening length using Eq.~\ref{yukawa}, at $z$ = 0.03 mm. ({\bf A}) In the 15-particle experiment, the interaction between small particles 1 and 2 ($s_1 = 0.234$ mm, $s_2 = 0.232$ mm, cyan) and between large particles 4 and 5 ($s_4 = -0.150$ mm, $s_5 = -0.161$ mm, gray) have a distinctly different decay with length scale $\lambda$. The solid lines are fits using Eq.~\ref{yukawa}. Note that a larger $\lambda$ means slower decay. ({\bf B}) The mass of all particles inferred from the drag coefficient ($m_\gamma$) versus the mass inferred from the particle interaction ($m_\text{int}$). Different colors represent the 5 different experiments (Table 1). The dashed line is the theoretical value of $m_\gamma = m_\text{int}$. The gray box represents particles with an average diameter of 12.8 $\pm$ 0.32 $\mu$m, corresponding to a mass of $m_0 = $1.65 $\pm$ 0.12 ng, which is necessary for quantifying the mass (see {\bf SI} for more information). ({\bf C}) Particles charge, $q$, versus $m_\text{int}$, both inferred from the fitting procedure using Eq.~\ref{yukawa}. The dashed lines are power law fits with the fitting power $p$ displayed alongside the lines. In both panels, the two clusters of purple and orange data (indicated by the arrows) each consist of 5 similar particles whose manufacturer-labeled diameters are 9.46 $\pm$ 0.10 $\mu$m (0.66 $\pm$ 0.02 ng) and 8.00 $\pm$ 0.09 $\mu$m (0.40 $\pm$ 0.01 ng), respectively. Inset: the fitting power $p$ versus the plasma pressure $P$. Note that the blue and green data coincide.}
\label{p4}
\end{figure*}

In addition to the dependence on $\rho$, the model can predict the dependence of the interaction force on $z$, revealing the spatial structure of the plasma sheath. Figure \hyperref[p3]{3\textit{C}} shows the reciprocal reduced force versus $z$ for particles 1 and 2 when $z_1=z_2=z$. At larger $z$, the force is nearly uniform, but then rises precipitously as $z$ decreases, more than a factor of two over a span of 200 $\mu$m. This sharp rise is mostly due to the variation of accumulated charge on each particle. In the bulk plasma, properties such as the ion and electron temperature and density are expected to be constant \cite{bohm1950effects, douglass2011dust}. Thus, the particle charge should also be constant. However, inside the plasma sheath, these properties change, and the charge on the particles can increase dramatically \cite{harper2020origin,douglass2012determination}. This is also evidenced by an increase of the screening length ($\lambda$) at the boundary of the plasma sheath (Fig.~S1). Additionally, we show the model's prediction of the reduced environmental force ($\vec{f}_i^\text{env}$) in Fig.~\hyperref[p3]{3\textit{D}}. 
This force acts on each particle separately, and is due to local electric fields and ion drag forces that trap the particle and drive its vortical motion, {\color{black}resulting in trajectories in the $xy$-plane that resemble Fig.~\hyperref[p1]{1\textit{D}}}. Taken together, Fig.~\hyperref[p3]{3} shows how our ML model can turn the particles into non-intrusive, local probes of the plasma environment.   

\subsection*{Estimating particle mass, charge, and interaction range}
\label{2c}
In many-body systems, measured properties of individual particles are often inaccessible or assumed from simple theories, yet our ML approach can infer both the mass and charge of each particle from experimental data alone. Using nonlinear regression (see \textit{Materials and Methods}), we simultaneously fitted the model's predicted interaction (e.g., Fig.~\hyperref[p3]{3\textit{B}}) to Eq.~\ref{yukawa} for every pair of particles in each experiment at $z = 0.03$ mm, with fitting parameters $m_i$, $q_i$, $q_j$, and $\lambda_{ij}$. To obtain good fits, it was necessary to allow the screening length ($\lambda_{ij}$) to vary between particle pairs, rather than be represented by a single constant that only depends on the plasma environment. This is evidenced in Fig.~\hyperref[p4]{4\textit{A}}, where $f_{ij}$ is plotted for a pair of small particles, and a pair of large particles. The screening length varies by almost a factor of 3. 

In the plasma sheath where particles are levitated, the supersonic motion of ions towards the electrode (negative $z$-direction) diminishes their ability to screen the charged particles  \cite{konopka1997central,ludwig2012wake}, meaning that, to the lowest order, $\lambda$ should be determined by the electron screening length (1-2 mm in our experiments \cite{harper2020origin}). However, the effective interactions between particles involve their associated ion wakes; the same wakes that give rise to non-reciprocal interactions (Fig.~\hyperref[p1]{1\textit{C}}). As the particle separation $\rho\rightarrow 0$, particles repel strongly through a Coulomb force representing the \emph{actual} charge on each particle. For large $\rho$, the effective particle charge is reduced by the virtual positive charge (ion wake). Thus, fitting the total interaction with Eq.~\ref{yukawa} should result in $\lambda$ being significantly less than the plasma Debye length. Also, $\lambda$ should depend on the strength and spatial extent of each particle's ion wake, which can lead to an apparent dependence on particle size. Indeed, an increase of $\lambda$ with particle size has been reported in experiments examining the linearized vibrational motion of dust particles \cite{carstensen2010determination}, and our results firmly demonstrate that Eq.~\ref{yukawa} is an approximation whose parameters must be carefully interpreted when considering effective particle interactions in dusty plasmas.


In addition to $\lambda$, our fitting procedure provides the mass and charge of each particle from the interaction. To validate this procedure, we obtained an independent estimate of the mass from the inferred damping coefficient ($\gamma$) by computing the particle's diameter using Eq.~\ref{epstein} and assuming the particles were spheres with density 1510 kg$\cdot$m$^{-3}$. These two independent masses, denoted $m_{i,\text{int}}$ and $m_{i,\gamma}$, show excellent agreement (Fig.~\hyperref[p4]{4\textit{B}}), demonstrating that the model correctly infers each term in Eq.~\ref{model} using experimental data. However, the inference of the particle charge ($q_i$) from Eq.~\ref{yukawa} reveals important discrepancies from widely-used theoretical assumptions. Orbital-motion-limited (OML) theory predicts the charge on a spherical particle in a dusty plasma under the assumption the electron and ion temperatures (and densities) are known and uniform, and collisions are ignored \cite{shukla2009colloquium,shukla2015introduction,ignatov2005basics, melzer2019physics}. Under these assumptions, two particles of different sizes \emph{should} act as spherical capacitors and have the same floating potential, $V_i=2\pi\epsilon_0 d_i q_i$. Thus, we expect $q_i\propto m_i^{1/3}$ since $m_i\propto d_i^3$. 
We tested this relationship by fitting the inferred charge versus mass in all 5 experiments using $q_i\propto m_i^{p}$. As shown in Fig.~\hyperref[p4]{4\textit{C}}, the power $p$ ranged from 0.30-0.80, and increased monotonically with pressure $P$ (Fig.~\hyperref[p4]{4\textit{C}} inset). Thus, even when the particle charge is inferred at the same $z$-position, where plasma properties should be the same for all particles, the power $p$ can vary substantially from the expected value of 1/3. 

The validation of the particle mass (Fig.~\hyperref[p4]{4\textit{B}}) suggests that the prefactor $A$ in Eq.~\ref{yukawa} is estimated correctly, and thus so is the particle charge. While it is possible that the presence of a positive ion wake reduces the effective charge of each particle at large $\rho$, this effect would be negligible as $\rho\rightarrow 0$, where the quality of the fits to the interaction force are equally good. Given the pressure dependence observed in the inset of Fig.~\hyperref[p4]{4\textit{C}}, it is natural to ascribe this variation in $p$ to collisions between ions and neutral atoms, which are often ignored in theories of particle charging, yet collisions should reduce the charge of larger particles due to their increased capture radius \cite{galli2010charging,gatti2008analytical}, thereby making $p<1/3$. As such, the origin of this discrepancy from common theoretical assumptions remains unclear, but our results highlight the need for more comprehensive theories of particle charging in plasma sheaths. Lastly, to ensure that our measurements of the screening length and particle charge are not artifacts of the inference process and accurately represent the physics, we simulated systems of many particles with similar non-reciprocal forces and environmental forces as in the experiment, and required that $q_i\propto m_i^{1/3}$ and a screening length independent of particle size (see the full details in \textbf{SI}, and Movie S8). The model achieved a validation $R^2$ = 0.9989, and showed a 35\% reduction of the inferred screening length in the presence of virtual positive charges representing the ion wakes beneath each particle (Fig.~S2A). Importantly, the model performed remarkably when extracting the reduced mass and charge of each particle (Fig.~S2B-C), demonstrating that the inferred deviations from $q\propto m^{1/3}$ in experimental data are likely real.

\section*{Summary and Conclusions}

We have developed a machine learning model that accurately infers the forces acting on individual particles in a many-body system. What makes this model different from past approaches is its ability to approximate complex, nonlinear interaction laws using NNs, 
treat particles as different individuals, build in physical symmetries into the model structure, and to learn purely from experimental data. By applying this new approach to dusty plasmas, we learned both environmental forces and pairwise interaction forces between particles, and extracted the mass and charge of each particle {\em in situ}. In doing so, we verified theoretical predictions of non-reciprocal and attractive forces between dust particles, discovered an unexpected dependence of the screening length on the size of interacting particles, and discovered unexpected deviations from OML theory (where $q\propto m^{1/3}$).


Furthermore, We highlight that the primary challenge in uncovering the dependence of $\lambda$ and $q$ on particle size lies in controlling the variable $z$, since $\lambda$ and $q$ steeply vary with the depth in the plasma sheath ($z$-direction). In equilibrium, heavier particles settle at lower $z$, rendering normal mode analysis approaches near the particles' equilibrium positions ineffective for comparing particles with different sizes \cite{ding2019nonlinear,ding2021machine,yu2022extracting,nunomura2002dispersion,couedel2009first,couedel2010direct,zhdanov2009mode}.
Therefore, we expect these results to serve as seeds for new directions of research in dusty plasma physics. 

Outside of dusty plasma research, our ML approach is widely applicable to physical and biological systems composed of many interacting agents. They can be active or passive, with arbitrarily complex interactions. Although intuition guides the underlying symmetries and expected structure of the model, the ability to surpass intuition and avoid biased assumptions is an essential first step in discovering new scientific laws from experiments.

\section*{Materials and Methods}
{\color{black}
\subsection*{Dusty plasma experiments}
In our experimental vacuum chamber, a disk-shaped aluminum electrode with a 15 cm diameter was driven with 13.56 MHz RF power (2-7 W) to generate a weakly-ionized plasma. The ion and electron density was $\approx$ 2--6 $\times$ 10$^{13}$ m$^{-3}$, as measured with a custom Langmuir probe \cite{harper2020origin}. Since the electron temperature (1--2 eV) was much higher than the ion and neutral gas temperature (0.04 eV), the electrode developed a negative bias voltage (-15 to -5 V). Particles were introduced into the plasma by mechanical agitation of a reservoir and developed a negative bias voltage as well, leading to an electrostatic repulsion from the electrode that levitated them $\approx$ 4.2 mm above the electrode surface. Most experiments confined 10-20 spherical melamine-formaldehyde (MF) particles (microParticles GMBH) in the plasma, and we purposefully used a combination of manufactured particles with labeled diameters of 12.8$\pm$0.32 $\mu$m, 9.46$\pm$0.10 $\mu$m, and 8.00$\pm$0.09 $\mu$m since our model is able to handle different particle sizes. A unique feature of our experiments was a cylindrical neodymium magnet with diameter 7.5 cm placed inside the electrode, resulting in a nonuniform magnetic field of strength $\approx0.04$ T where the particles levitated. The gradient in the field produced a vortical ion flow and ion drag force on each particle, resulting in a highly-dynamic system of particles with circulation.

\subsection*{3D particle tracking method}

To track the position of each particle, we used a rapidly-scanning laser sheet synchronized to a high-speed camera, as shown in Fig.~\hyperref[p1]{1A}. The laser scanning frequency was 200 Hz with a peak-to-peak amplitude of $\approx$ 4 mm at the position of the particles. The high speed camera was set to record at 8000 fps, thus we obtained $\approx$ 40 images in one vertical sweep of the laser sheet. The resulting images were analyzed by TrackPy \cite{allan_2024_12708864}, and special care was taken to distinguish particles with very small separations. The ultimate spatial resolution of the in-plane $x$ and $y$ positions was 50 $\mu$m, and the vertical $z$ resolution was 100 $\mu$m. More complete details of the 3D tracking method can be found in references \cite{yu20233d,yu2022extracting}.   

\subsection*{Fitting of charge and mass for each particle}

For extracting $m_i$ and $q_i$, we used a screened Coulomb interaction $f^C$: 
\begin{align}
    &f^C(\rho; q_i, q_j, m_i, \lambda_i, \lambda_j) = \\
    &\frac{q_iq_j}{4\pi\epsilon_0 m_i\rho}\left(\frac{1}{\rho} + \frac{1}{\sqrt{\lambda_i\lambda_j}}\right)e^{-\rho/\sqrt{\lambda_i\lambda_j}}.
    \label{fiteq}
\end{align}
In order to find the mass and charge of all particles at a specific $z$ position, we performed a global least-squares fit of every pair of particle interactions. For example, for a given $z$ position, let $\Bar{f}_{ij}(\rho)$ represents the model's prediction of particle $j$'s reduced force on $i$ at vertical position $z_i=z_j=z$ and horizontal separation $\rho$: 
\begin{equation}
   \Bar{f}_{ij}(\rho) = \frac{g_{\text{int}}(\rho, z, z, s_i, s_j)}{\rho}.
\end{equation}
In the fitting procedure, we aim at finding the optimal values of $\{q_i, q_j, m_i, \lambda_i, \lambda_j\}$ that minimize the following loss function:
\begin{equation}
    L^C = \sum_{i = 0}^{N_p}\sum_{j = 0, j\neq i}^{N_p}\sum_{\rho}^{[a, b, c]}\left(\Bar{f}_{ij}(\rho) - f^C(\rho;q_i,q_j,m_i,\lambda_i,\lambda_j)\right)^2.
\end{equation}
Here $[a,b,c]$ defines which particle interactions to include in the sum. The minimum separation is  $\rho=a$, the maximum separation is $\rho=b$, and particles within a small range $c$ are included at each separation. For Fig.~\hyperref[p3]{3}, we chose $a$ = 0.3 mm, $b$ = 1.2 mm, and $c$ = 0.01 mm. 
We note that the charge and the mass are coupled in the fitting procedure since they appear as a ratio. For example, if we decrease all particles' mass by a factor of 4, and decrease all particle's charges by a factor of 2, the fitting quality wouldn't change. Thus, we added a constraint in the fitting that the average mass of the particles in the shaded area in Fig.~\hyperref[p4]{4\textit{B}} should be 1.65 ng, the average mass reported by the manufacturer. The above procedure was implemented for each of the 10 trained models, and the average $q_i$ and $m_i$ over all 10 models plus their standard deviation is reported in Fig.~\hyperref[p4]{4}.

}


\begin{acknowledgements}
\textbf{Funding:} This material is based upon work supported by the National Science Foundation under Grant No.\ 2010524, the U.S. Department of Energy, Office of Science, Office of Fusion Energy Sciences program under Award No.\ DESC0021290, and by the Simons Foundation Investigators Program. \noindent\textbf{Author contributions:} WY and JCB conducted experimental research. WY, JCB, EA, and IN developed the machine learning theory. WY wrote all code for the model, and EA reviewed the code. All authors wrote and reviewed the manuscript. \noindent\textbf{Competing interests:} The authors declare no competing interests.
\end{acknowledgements}

\bibliography{Dusty_Plasma_Arxiv}

\begin{thebibliography}{85}%
\makeatletter
\providecommand \@ifxundefined [1]{%
 \@ifx{#1\undefined}
}%
\providecommand \@ifnum [1]{%
 \ifnum #1\expandafter \@firstoftwo
 \else \expandafter \@secondoftwo
 \fi
}%
\providecommand \@ifx [1]{%
 \ifx #1\expandafter \@firstoftwo
 \else \expandafter \@secondoftwo
 \fi
}%
\providecommand \natexlab [1]{#1}%
\providecommand \enquote  [1]{``#1''}%
\providecommand \bibnamefont  [1]{#1}%
\providecommand \bibfnamefont [1]{#1}%
\providecommand \citenamefont [1]{#1}%
\providecommand \href@noop [0]{\@secondoftwo}%
\providecommand \href [0]{\begingroup \@sanitize@url \@href}%
\providecommand \@href[1]{\@@startlink{#1}\@@href}%
\providecommand \@@href[1]{\endgroup#1\@@endlink}%
\providecommand \@sanitize@url [0]{\catcode `\\12\catcode `\$12\catcode `\&12\catcode `\#12\catcode `\^12\catcode `\_12\catcode `\%12\relax}%
\providecommand \@@startlink[1]{}%
\providecommand \@@endlink[0]{}%
\providecommand \url  [0]{\begingroup\@sanitize@url \@url }%
\providecommand \@url [1]{\endgroup\@href {#1}{\urlprefix }}%
\providecommand \urlprefix  [0]{URL }%
\providecommand \Eprint [0]{\href }%
\providecommand \doibase [0]{http://dx.doi.org/}%
\providecommand \selectlanguage [0]{\@gobble}%
\providecommand \bibinfo  [0]{\@secondoftwo}%
\providecommand \bibfield  [0]{\@secondoftwo}%
\providecommand \translation [1]{[#1]}%
\providecommand \BibitemOpen [0]{}%
\providecommand \bibitemStop [0]{}%
\providecommand \bibitemNoStop [0]{.\EOS\space}%
\providecommand \EOS [0]{\spacefactor3000\relax}%
\providecommand \BibitemShut  [1]{\csname bibitem#1\endcsname}%
\let\auto@bib@innerbib\@empty
\bibitem [{\citenamefont {Merlino}(2021)}]{merlino2021dusty}%
  \BibitemOpen
  \bibfield  {author} {\bibinfo {author} {\bibfnamefont {R.}~\bibnamefont {Merlino}},\ }\href@noop {} {\bibfield  {journal} {\bibinfo  {journal} {Advances in Physics: X}\ }\textbf {\bibinfo {volume} {6}},\ \bibinfo {pages} {1873859} (\bibinfo {year} {2021})}\BibitemShut {NoStop}%
\bibitem [{\citenamefont {Carleo}\ \emph {et~al.}(2019)\citenamefont {Carleo}, \citenamefont {Cirac}, \citenamefont {Cranmer}, \citenamefont {Daudet}, \citenamefont {Schuld}, \citenamefont {Tishby}, \citenamefont {Vogt-Maranto},\ and\ \citenamefont {Zdeborov{\'a}}}]{carleo2019machine}%
  \BibitemOpen
  \bibfield  {author} {\bibinfo {author} {\bibfnamefont {G.}~\bibnamefont {Carleo}}, \bibinfo {author} {\bibfnamefont {I.}~\bibnamefont {Cirac}}, \bibinfo {author} {\bibfnamefont {K.}~\bibnamefont {Cranmer}}, \bibinfo {author} {\bibfnamefont {L.}~\bibnamefont {Daudet}}, \bibinfo {author} {\bibfnamefont {M.}~\bibnamefont {Schuld}}, \bibinfo {author} {\bibfnamefont {N.}~\bibnamefont {Tishby}}, \bibinfo {author} {\bibfnamefont {L.}~\bibnamefont {Vogt-Maranto}}, \ and\ \bibinfo {author} {\bibfnamefont {L.}~\bibnamefont {Zdeborov{\'a}}},\ }\href@noop {} {\bibfield  {journal} {\bibinfo  {journal} {Reviews of Modern Physics}\ }\textbf {\bibinfo {volume} {91}},\ \bibinfo {pages} {045002} (\bibinfo {year} {2019})}\BibitemShut {NoStop}%
\bibitem [{\citenamefont {Pineda}\ \emph {et~al.}(2023)\citenamefont {Pineda}, \citenamefont {Midtvedt}, \citenamefont {Bachimanchi}, \citenamefont {No{\'e}}, \citenamefont {Midtvedt}, \citenamefont {Volpe},\ and\ \citenamefont {Manzo}}]{pineda2023geometric}%
  \BibitemOpen
  \bibfield  {author} {\bibinfo {author} {\bibfnamefont {J.}~\bibnamefont {Pineda}}, \bibinfo {author} {\bibfnamefont {B.}~\bibnamefont {Midtvedt}}, \bibinfo {author} {\bibfnamefont {H.}~\bibnamefont {Bachimanchi}}, \bibinfo {author} {\bibfnamefont {S.}~\bibnamefont {No{\'e}}}, \bibinfo {author} {\bibfnamefont {D.}~\bibnamefont {Midtvedt}}, \bibinfo {author} {\bibfnamefont {G.}~\bibnamefont {Volpe}}, \ and\ \bibinfo {author} {\bibfnamefont {C.}~\bibnamefont {Manzo}},\ }\href@noop {} {\bibfield  {journal} {\bibinfo  {journal} {Nature Machine Intelligence}\ }\textbf {\bibinfo {volume} {5}},\ \bibinfo {pages} {71} (\bibinfo {year} {2023})}\BibitemShut {NoStop}%
\bibitem [{\citenamefont {Br{\"u}ckner}\ \emph {et~al.}(2021)\citenamefont {Br{\"u}ckner}, \citenamefont {Arlt}, \citenamefont {Fink}, \citenamefont {Ronceray}, \citenamefont {R{\"a}dler},\ and\ \citenamefont {Broedersz}}]{bruckner2021learning}%
  \BibitemOpen
  \bibfield  {author} {\bibinfo {author} {\bibfnamefont {D.~B.}\ \bibnamefont {Br{\"u}ckner}}, \bibinfo {author} {\bibfnamefont {N.}~\bibnamefont {Arlt}}, \bibinfo {author} {\bibfnamefont {A.}~\bibnamefont {Fink}}, \bibinfo {author} {\bibfnamefont {P.}~\bibnamefont {Ronceray}}, \bibinfo {author} {\bibfnamefont {J.~O.}\ \bibnamefont {R{\"a}dler}}, \ and\ \bibinfo {author} {\bibfnamefont {C.~P.}\ \bibnamefont {Broedersz}},\ }\href@noop {} {\bibfield  {journal} {\bibinfo  {journal} {Proceedings of the National Academy of Sciences}\ }\textbf {\bibinfo {volume} {118}},\ \bibinfo {pages} {e2016602118} (\bibinfo {year} {2021})}\BibitemShut {NoStop}%
\bibitem [{\citenamefont {Toner}\ and\ \citenamefont {Tu}(1998)}]{toner1998flocks}%
  \BibitemOpen
  \bibfield  {author} {\bibinfo {author} {\bibfnamefont {J.}~\bibnamefont {Toner}}\ and\ \bibinfo {author} {\bibfnamefont {Y.}~\bibnamefont {Tu}},\ }\href@noop {} {\bibfield  {journal} {\bibinfo  {journal} {Physical review E}\ }\textbf {\bibinfo {volume} {58}},\ \bibinfo {pages} {4828} (\bibinfo {year} {1998})}\BibitemShut {NoStop}%
\bibitem [{\citenamefont {Melzer}\ \emph {et~al.}(2019{\natexlab{a}})\citenamefont {Melzer} \emph {et~al.}}]{melzer2019physics}%
  \BibitemOpen
  \bibfield  {author} {\bibinfo {author} {\bibfnamefont {A.}~\bibnamefont {Melzer}} \emph {et~al.},\ }\href@noop {} {\emph {\bibinfo {title} {Physics of dusty plasmas}}},\ Vol.\ \bibinfo {volume} {962}\ (\bibinfo  {publisher} {Springer},\ \bibinfo {year} {2019})\BibitemShut {NoStop}%
\bibitem [{\citenamefont {Hippler}\ \emph {et~al.}(2008)\citenamefont {Hippler}, \citenamefont {Kersten}, \citenamefont {Schmidt},\ and\ \citenamefont {Schoenbach}}]{hippler2008low}%
  \BibitemOpen
  \bibfield  {author} {\bibinfo {author} {\bibfnamefont {R.}~\bibnamefont {Hippler}}, \bibinfo {author} {\bibfnamefont {H.}~\bibnamefont {Kersten}}, \bibinfo {author} {\bibfnamefont {M.}~\bibnamefont {Schmidt}}, \ and\ \bibinfo {author} {\bibfnamefont {K.~H.}\ \bibnamefont {Schoenbach}},\ }\href@noop {} {\bibfield  {journal} {\bibinfo  {journal} {Eds R Hippler et al, Berlin: Wiley}\ }\textbf {\bibinfo {volume} {787}} (\bibinfo {year} {2008})}\BibitemShut {NoStop}%
\bibitem [{\citenamefont {Chaudhuri}\ \emph {et~al.}(2011)\citenamefont {Chaudhuri}, \citenamefont {Ivlev}, \citenamefont {Khrapak}, \citenamefont {Thomas},\ and\ \citenamefont {Morfill}}]{chaudhuri2011complex}%
  \BibitemOpen
  \bibfield  {author} {\bibinfo {author} {\bibfnamefont {M.}~\bibnamefont {Chaudhuri}}, \bibinfo {author} {\bibfnamefont {A.~V.}\ \bibnamefont {Ivlev}}, \bibinfo {author} {\bibfnamefont {S.~A.}\ \bibnamefont {Khrapak}}, \bibinfo {author} {\bibfnamefont {H.~M.}\ \bibnamefont {Thomas}}, \ and\ \bibinfo {author} {\bibfnamefont {G.~E.}\ \bibnamefont {Morfill}},\ }\href@noop {} {\bibfield  {journal} {\bibinfo  {journal} {Soft Matter}\ }\textbf {\bibinfo {volume} {7}},\ \bibinfo {pages} {1287} (\bibinfo {year} {2011})}\BibitemShut {NoStop}%
\bibitem [{\citenamefont {Goertz}(1989)}]{goertz1989dusty}%
  \BibitemOpen
  \bibfield  {author} {\bibinfo {author} {\bibfnamefont {C.}~\bibnamefont {Goertz}},\ }\href@noop {} {\bibfield  {journal} {\bibinfo  {journal} {Reviews of Geophysics}\ }\textbf {\bibinfo {volume} {27}},\ \bibinfo {pages} {271} (\bibinfo {year} {1989})}\BibitemShut {NoStop}%
\bibitem [{\citenamefont {Wahlund}\ \emph {et~al.}(2009)\citenamefont {Wahlund}, \citenamefont {Andr{\'e}}, \citenamefont {Eriksson}, \citenamefont {Lundberg}, \citenamefont {Morooka}, \citenamefont {Shafiq}, \citenamefont {Averkamp}, \citenamefont {Gurnett}, \citenamefont {Hospodarsky}, \citenamefont {Kurth} \emph {et~al.}}]{wahlund2009detection}%
  \BibitemOpen
  \bibfield  {author} {\bibinfo {author} {\bibfnamefont {J.-E.}\ \bibnamefont {Wahlund}}, \bibinfo {author} {\bibfnamefont {M.}~\bibnamefont {Andr{\'e}}}, \bibinfo {author} {\bibfnamefont {A.}~\bibnamefont {Eriksson}}, \bibinfo {author} {\bibfnamefont {M.}~\bibnamefont {Lundberg}}, \bibinfo {author} {\bibfnamefont {M.}~\bibnamefont {Morooka}}, \bibinfo {author} {\bibfnamefont {M.}~\bibnamefont {Shafiq}}, \bibinfo {author} {\bibfnamefont {T.}~\bibnamefont {Averkamp}}, \bibinfo {author} {\bibfnamefont {D.}~\bibnamefont {Gurnett}}, \bibinfo {author} {\bibfnamefont {G.}~\bibnamefont {Hospodarsky}}, \bibinfo {author} {\bibfnamefont {W.}~\bibnamefont {Kurth}},  \emph {et~al.},\ }\href@noop {} {\bibfield  {journal} {\bibinfo  {journal} {Planetary and Space Science}\ }\textbf {\bibinfo {volume} {57}},\ \bibinfo {pages} {1795} (\bibinfo {year} {2009})}\BibitemShut {NoStop}%
\bibitem [{\citenamefont {Shukla}(2002)}]{shukla2002dust}%
  \BibitemOpen
  \bibfield  {author} {\bibinfo {author} {\bibfnamefont {P.~K.}\ \bibnamefont {Shukla}},\ }\href@noop {} {\emph {\bibinfo {title} {Dust Plasma Interaction in Space}}}\ (\bibinfo  {publisher} {Nova Publishers},\ \bibinfo {year} {2002})\BibitemShut {NoStop}%
\bibitem [{\citenamefont {Kortshagen}(2009)}]{kortshagen2009nonthermal}%
  \BibitemOpen
  \bibfield  {author} {\bibinfo {author} {\bibfnamefont {U.}~\bibnamefont {Kortshagen}},\ }\href@noop {} {\bibfield  {journal} {\bibinfo  {journal} {Journal of Physics D: Applied Physics}\ }\textbf {\bibinfo {volume} {42}},\ \bibinfo {pages} {113001} (\bibinfo {year} {2009})}\BibitemShut {NoStop}%
\bibitem [{\citenamefont {Merlino}\ and\ \citenamefont {Goree}(2004)}]{merlino2004dusty}%
  \BibitemOpen
  \bibfield  {author} {\bibinfo {author} {\bibfnamefont {R.~L.}\ \bibnamefont {Merlino}}\ and\ \bibinfo {author} {\bibfnamefont {J.~A.}\ \bibnamefont {Goree}},\ }\href@noop {} {\bibfield  {journal} {\bibinfo  {journal} {Physics Today}\ }\textbf {\bibinfo {volume} {57}},\ \bibinfo {pages} {32} (\bibinfo {year} {2004})}\BibitemShut {NoStop}%
\bibitem [{\citenamefont {Beckers}\ \emph {et~al.}(2019)\citenamefont {Beckers}, \citenamefont {van~de Ven}, \citenamefont {van~der Horst}, \citenamefont {Astakhov},\ and\ \citenamefont {Banine}}]{beckers2019euv}%
  \BibitemOpen
  \bibfield  {author} {\bibinfo {author} {\bibfnamefont {J.}~\bibnamefont {Beckers}}, \bibinfo {author} {\bibfnamefont {T.}~\bibnamefont {van~de Ven}}, \bibinfo {author} {\bibfnamefont {R.}~\bibnamefont {van~der Horst}}, \bibinfo {author} {\bibfnamefont {D.}~\bibnamefont {Astakhov}}, \ and\ \bibinfo {author} {\bibfnamefont {V.}~\bibnamefont {Banine}},\ }\href@noop {} {\bibfield  {journal} {\bibinfo  {journal} {Applied Sciences}\ }\textbf {\bibinfo {volume} {9}},\ \bibinfo {pages} {2827} (\bibinfo {year} {2019})}\BibitemShut {NoStop}%
\bibitem [{\citenamefont {Winter}(2000)}]{winter2000dust}%
  \BibitemOpen
  \bibfield  {author} {\bibinfo {author} {\bibfnamefont {J.}~\bibnamefont {Winter}},\ }\href@noop {} {\bibfield  {journal} {\bibinfo  {journal} {Physics of Plasmas}\ }\textbf {\bibinfo {volume} {7}},\ \bibinfo {pages} {3862} (\bibinfo {year} {2000})}\BibitemShut {NoStop}%
\bibitem [{\citenamefont {Tsytovich}\ \emph {et~al.}(2007)\citenamefont {Tsytovich}, \citenamefont {Morfill}, \citenamefont {Fortov}, \citenamefont {Gusein-Zade}, \citenamefont {Klumov},\ and\ \citenamefont {Vladimirov}}]{tsytovich2007plasma}%
  \BibitemOpen
  \bibfield  {author} {\bibinfo {author} {\bibfnamefont {V.}~\bibnamefont {Tsytovich}}, \bibinfo {author} {\bibfnamefont {G.}~\bibnamefont {Morfill}}, \bibinfo {author} {\bibfnamefont {V.}~\bibnamefont {Fortov}}, \bibinfo {author} {\bibfnamefont {N.}~\bibnamefont {Gusein-Zade}}, \bibinfo {author} {\bibfnamefont {B.}~\bibnamefont {Klumov}}, \ and\ \bibinfo {author} {\bibfnamefont {S.}~\bibnamefont {Vladimirov}},\ }\href@noop {} {\bibfield  {journal} {\bibinfo  {journal} {New Journal of Physics}\ }\textbf {\bibinfo {volume} {9}},\ \bibinfo {pages} {263} (\bibinfo {year} {2007})}\BibitemShut {NoStop}%
\bibitem [{\citenamefont {Matthews}\ \emph {et~al.}(2020)\citenamefont {Matthews}, \citenamefont {Sanford}, \citenamefont {Kostadinova}, \citenamefont {Ashrafi}, \citenamefont {Guay},\ and\ \citenamefont {Hyde}}]{matthews2020dust}%
  \BibitemOpen
  \bibfield  {author} {\bibinfo {author} {\bibfnamefont {L.~S.}\ \bibnamefont {Matthews}}, \bibinfo {author} {\bibfnamefont {D.~L.}\ \bibnamefont {Sanford}}, \bibinfo {author} {\bibfnamefont {E.~G.}\ \bibnamefont {Kostadinova}}, \bibinfo {author} {\bibfnamefont {K.~S.}\ \bibnamefont {Ashrafi}}, \bibinfo {author} {\bibfnamefont {E.}~\bibnamefont {Guay}}, \ and\ \bibinfo {author} {\bibfnamefont {T.~W.}\ \bibnamefont {Hyde}},\ }\href@noop {} {\bibfield  {journal} {\bibinfo  {journal} {Physics of Plasmas}\ }\textbf {\bibinfo {volume} {27}} (\bibinfo {year} {2020})}\BibitemShut {NoStop}%
\bibitem [{\citenamefont {Melzer}\ \emph {et~al.}(2021)\citenamefont {Melzer}, \citenamefont {Kr{\"u}ger}, \citenamefont {Maier},\ and\ \citenamefont {Sch{\"u}tt}}]{melzer2021physics}%
  \BibitemOpen
  \bibfield  {author} {\bibinfo {author} {\bibfnamefont {A.}~\bibnamefont {Melzer}}, \bibinfo {author} {\bibfnamefont {H.}~\bibnamefont {Kr{\"u}ger}}, \bibinfo {author} {\bibfnamefont {D.}~\bibnamefont {Maier}}, \ and\ \bibinfo {author} {\bibfnamefont {S.}~\bibnamefont {Sch{\"u}tt}},\ }\href@noop {} {\bibfield  {journal} {\bibinfo  {journal} {Reviews of Modern Plasma Physics}\ }\textbf {\bibinfo {volume} {5}},\ \bibinfo {pages} {11} (\bibinfo {year} {2021})}\BibitemShut {NoStop}%
\bibitem [{\citenamefont {Thomas}\ \emph {et~al.}(2012)\citenamefont {Thomas}, \citenamefont {Merlino},\ and\ \citenamefont {Rosenberg}}]{thomas2012magnetized}%
  \BibitemOpen
  \bibfield  {author} {\bibinfo {author} {\bibfnamefont {E.}~\bibnamefont {Thomas}}, \bibinfo {author} {\bibfnamefont {R.}~\bibnamefont {Merlino}}, \ and\ \bibinfo {author} {\bibfnamefont {M.}~\bibnamefont {Rosenberg}},\ }\href@noop {} {\bibfield  {journal} {\bibinfo  {journal} {Plasma Physics and Controlled Fusion}\ }\textbf {\bibinfo {volume} {54}},\ \bibinfo {pages} {124034} (\bibinfo {year} {2012})}\BibitemShut {NoStop}%
\bibitem [{\citenamefont {Melzer}\ and\ \citenamefont {Goree}(2008)}]{melzer2008fundamentals}%
  \BibitemOpen
  \bibfield  {author} {\bibinfo {author} {\bibfnamefont {A.}~\bibnamefont {Melzer}}\ and\ \bibinfo {author} {\bibfnamefont {J.}~\bibnamefont {Goree}},\ }in\ \href@noop {} {\emph {\bibinfo {booktitle} {Low temperature plasmas fundamentals, technologies, and techniques}}},\ Vol.~\bibinfo {volume} {1},\ \bibinfo {editor} {edited by\ \bibinfo {editor} {\bibfnamefont {R.}~\bibnamefont {Hippler}}, \bibinfo {editor} {\bibfnamefont {H.}~\bibnamefont {Kersten}}, \bibinfo {editor} {\bibfnamefont {M.}~\bibnamefont {Schmidt}}, \ and\ \bibinfo {editor} {\bibfnamefont {K.~H.}\ \bibnamefont {Schoenbach}}}\ (\bibinfo  {publisher} {Wiley-VCH},\ \bibinfo {year} {2008})\ \bibinfo {edition} {2nd}\ ed.,\ pp.\ \bibinfo {pages} {157--206}\BibitemShut {NoStop}%
\bibitem [{\citenamefont {Melzer}\ \emph {et~al.}(2019{\natexlab{b}})\citenamefont {Melzer}, \citenamefont {Krueger}, \citenamefont {Schuett},\ and\ \citenamefont {Mulsow}}]{melzer2019finite}%
  \BibitemOpen
  \bibfield  {author} {\bibinfo {author} {\bibfnamefont {A.}~\bibnamefont {Melzer}}, \bibinfo {author} {\bibfnamefont {H.}~\bibnamefont {Krueger}}, \bibinfo {author} {\bibfnamefont {S.}~\bibnamefont {Schuett}}, \ and\ \bibinfo {author} {\bibfnamefont {M.}~\bibnamefont {Mulsow}},\ }\href@noop {} {\bibfield  {journal} {\bibinfo  {journal} {Physics of Plasmas}\ }\textbf {\bibinfo {volume} {26}},\ \bibinfo {pages} {093702} (\bibinfo {year} {2019}{\natexlab{b}})}\BibitemShut {NoStop}%
\bibitem [{\citenamefont {Nikolaev}\ and\ \citenamefont {Timofeev}(2021)}]{nikolaev2021nonhomogeneity}%
  \BibitemOpen
  \bibfield  {author} {\bibinfo {author} {\bibfnamefont {V.}~\bibnamefont {Nikolaev}}\ and\ \bibinfo {author} {\bibfnamefont {A.}~\bibnamefont {Timofeev}},\ }\href@noop {} {\bibfield  {journal} {\bibinfo  {journal} {Physics of Plasmas}\ }\textbf {\bibinfo {volume} {28}},\ \bibinfo {pages} {033704} (\bibinfo {year} {2021})}\BibitemShut {NoStop}%
\bibitem [{\citenamefont {Kolotinskii}\ \emph {et~al.}(2021)\citenamefont {Kolotinskii}, \citenamefont {Nikolaev},\ and\ \citenamefont {Timofeev}}]{kolotinskii2021effect}%
  \BibitemOpen
  \bibfield  {author} {\bibinfo {author} {\bibfnamefont {D.~A.}\ \bibnamefont {Kolotinskii}}, \bibinfo {author} {\bibfnamefont {V.~S.}\ \bibnamefont {Nikolaev}}, \ and\ \bibinfo {author} {\bibfnamefont {A.~V.}\ \bibnamefont {Timofeev}},\ }\href@noop {} {\bibfield  {journal} {\bibinfo  {journal} {JETP Letters}\ }\textbf {\bibinfo {volume} {113}},\ \bibinfo {pages} {510} (\bibinfo {year} {2021})}\BibitemShut {NoStop}%
\bibitem [{\citenamefont {Vaulina}\ \emph {et~al.}(2015)\citenamefont {Vaulina}, \citenamefont {Lisina},\ and\ \citenamefont {Lisin}}]{vaulina2015energy}%
  \BibitemOpen
  \bibfield  {author} {\bibinfo {author} {\bibfnamefont {O.}~\bibnamefont {Vaulina}}, \bibinfo {author} {\bibfnamefont {I.}~\bibnamefont {Lisina}}, \ and\ \bibinfo {author} {\bibfnamefont {E.}~\bibnamefont {Lisin}},\ }\href@noop {} {\bibfield  {journal} {\bibinfo  {journal} {Journal of Experimental and Theoretical Physics}\ }\textbf {\bibinfo {volume} {121}},\ \bibinfo {pages} {717} (\bibinfo {year} {2015})}\BibitemShut {NoStop}%
\bibitem [{\citenamefont {Ivlev}\ \emph {et~al.}(2015)\citenamefont {Ivlev}, \citenamefont {Bartnick}, \citenamefont {Heinen}, \citenamefont {Du}, \citenamefont {Nosenko},\ and\ \citenamefont {L{\"o}wen}}]{ivlev2015statistical}%
  \BibitemOpen
  \bibfield  {author} {\bibinfo {author} {\bibfnamefont {A.~V.}\ \bibnamefont {Ivlev}}, \bibinfo {author} {\bibfnamefont {J.}~\bibnamefont {Bartnick}}, \bibinfo {author} {\bibfnamefont {M.}~\bibnamefont {Heinen}}, \bibinfo {author} {\bibfnamefont {C.-R.}\ \bibnamefont {Du}}, \bibinfo {author} {\bibfnamefont {V.}~\bibnamefont {Nosenko}}, \ and\ \bibinfo {author} {\bibfnamefont {H.}~\bibnamefont {L{\"o}wen}},\ }\href@noop {} {\bibfield  {journal} {\bibinfo  {journal} {Physical Review X}\ }\textbf {\bibinfo {volume} {5}},\ \bibinfo {pages} {011035} (\bibinfo {year} {2015})}\BibitemShut {NoStop}%
\bibitem [{\citenamefont {Ding}\ \emph {et~al.}(2019)\citenamefont {Ding}, \citenamefont {Qiao}, \citenamefont {Kong}, \citenamefont {Matthews},\ and\ \citenamefont {Hyde}}]{ding2019nonlinear}%
  \BibitemOpen
  \bibfield  {author} {\bibinfo {author} {\bibfnamefont {Z.}~\bibnamefont {Ding}}, \bibinfo {author} {\bibfnamefont {K.}~\bibnamefont {Qiao}}, \bibinfo {author} {\bibfnamefont {J.}~\bibnamefont {Kong}}, \bibinfo {author} {\bibfnamefont {L.~S.}\ \bibnamefont {Matthews}}, \ and\ \bibinfo {author} {\bibfnamefont {T.~W.}\ \bibnamefont {Hyde}},\ }\href@noop {} {\bibfield  {journal} {\bibinfo  {journal} {Plasma Physics and Controlled Fusion}\ }\textbf {\bibinfo {volume} {61}},\ \bibinfo {pages} {055004} (\bibinfo {year} {2019})}\BibitemShut {NoStop}%
\bibitem [{\citenamefont {Mukhopadhyay}\ and\ \citenamefont {Goree}(2012)}]{mukhopadhyay2012two}%
  \BibitemOpen
  \bibfield  {author} {\bibinfo {author} {\bibfnamefont {A.~K.}\ \bibnamefont {Mukhopadhyay}}\ and\ \bibinfo {author} {\bibfnamefont {J.}~\bibnamefont {Goree}},\ }\href@noop {} {\bibfield  {journal} {\bibinfo  {journal} {Physical review letters}\ }\textbf {\bibinfo {volume} {109}},\ \bibinfo {pages} {165003} (\bibinfo {year} {2012})}\BibitemShut {NoStop}%
\bibitem [{\citenamefont {Ding}\ \emph {et~al.}(2021)\citenamefont {Ding}, \citenamefont {Matthews},\ and\ \citenamefont {Hyde}}]{ding2021machine}%
  \BibitemOpen
  \bibfield  {author} {\bibinfo {author} {\bibfnamefont {Z.}~\bibnamefont {Ding}}, \bibinfo {author} {\bibfnamefont {L.~S.}\ \bibnamefont {Matthews}}, \ and\ \bibinfo {author} {\bibfnamefont {T.~W.}\ \bibnamefont {Hyde}},\ }\href@noop {} {\bibfield  {journal} {\bibinfo  {journal} {Machine Learning: Science and Technology}\ }\textbf {\bibinfo {volume} {2}},\ \bibinfo {pages} {035017} (\bibinfo {year} {2021})}\BibitemShut {NoStop}%
\bibitem [{\citenamefont {Yu}\ \emph {et~al.}(2022)\citenamefont {Yu}, \citenamefont {Cho},\ and\ \citenamefont {Burton}}]{yu2022extracting}%
  \BibitemOpen
  \bibfield  {author} {\bibinfo {author} {\bibfnamefont {W.}~\bibnamefont {Yu}}, \bibinfo {author} {\bibfnamefont {J.}~\bibnamefont {Cho}}, \ and\ \bibinfo {author} {\bibfnamefont {J.~C.}\ \bibnamefont {Burton}},\ }\href@noop {} {\bibfield  {journal} {\bibinfo  {journal} {Physical Review E}\ }\textbf {\bibinfo {volume} {106}},\ \bibinfo {pages} {035303} (\bibinfo {year} {2022})}\BibitemShut {NoStop}%
\bibitem [{\citenamefont {Nunomura}\ \emph {et~al.}(2002)\citenamefont {Nunomura}, \citenamefont {Goree}, \citenamefont {Hu}, \citenamefont {Wang},\ and\ \citenamefont {Bhattacharjee}}]{nunomura2002dispersion}%
  \BibitemOpen
  \bibfield  {author} {\bibinfo {author} {\bibfnamefont {S.}~\bibnamefont {Nunomura}}, \bibinfo {author} {\bibfnamefont {J.}~\bibnamefont {Goree}}, \bibinfo {author} {\bibfnamefont {S.}~\bibnamefont {Hu}}, \bibinfo {author} {\bibfnamefont {X.}~\bibnamefont {Wang}}, \ and\ \bibinfo {author} {\bibfnamefont {A.}~\bibnamefont {Bhattacharjee}},\ }\href@noop {} {\bibfield  {journal} {\bibinfo  {journal} {Physical Review E}\ }\textbf {\bibinfo {volume} {65}},\ \bibinfo {pages} {066402} (\bibinfo {year} {2002})}\BibitemShut {NoStop}%
\bibitem [{\citenamefont {Cou{\"e}del}\ \emph {et~al.}(2009)\citenamefont {Cou{\"e}del}, \citenamefont {Nosenko}, \citenamefont {Zhdanov}, \citenamefont {Ivlev}, \citenamefont {Thomas},\ and\ \citenamefont {Morfill}}]{couedel2009first}%
  \BibitemOpen
  \bibfield  {author} {\bibinfo {author} {\bibfnamefont {L.}~\bibnamefont {Cou{\"e}del}}, \bibinfo {author} {\bibfnamefont {V.}~\bibnamefont {Nosenko}}, \bibinfo {author} {\bibfnamefont {S.}~\bibnamefont {Zhdanov}}, \bibinfo {author} {\bibfnamefont {A.}~\bibnamefont {Ivlev}}, \bibinfo {author} {\bibfnamefont {H.}~\bibnamefont {Thomas}}, \ and\ \bibinfo {author} {\bibfnamefont {G.}~\bibnamefont {Morfill}},\ }\href@noop {} {\bibfield  {journal} {\bibinfo  {journal} {Physical review letters}\ }\textbf {\bibinfo {volume} {103}},\ \bibinfo {pages} {215001} (\bibinfo {year} {2009})}\BibitemShut {NoStop}%
\bibitem [{\citenamefont {Cou{\"e}del}\ \emph {et~al.}(2010)\citenamefont {Cou{\"e}del}, \citenamefont {Nosenko}, \citenamefont {Ivlev}, \citenamefont {Zhdanov}, \citenamefont {Thomas},\ and\ \citenamefont {Morfill}}]{couedel2010direct}%
  \BibitemOpen
  \bibfield  {author} {\bibinfo {author} {\bibfnamefont {L.}~\bibnamefont {Cou{\"e}del}}, \bibinfo {author} {\bibfnamefont {V.}~\bibnamefont {Nosenko}}, \bibinfo {author} {\bibfnamefont {A.}~\bibnamefont {Ivlev}}, \bibinfo {author} {\bibfnamefont {S.}~\bibnamefont {Zhdanov}}, \bibinfo {author} {\bibfnamefont {H.}~\bibnamefont {Thomas}}, \ and\ \bibinfo {author} {\bibfnamefont {G.}~\bibnamefont {Morfill}},\ }\href@noop {} {\bibfield  {journal} {\bibinfo  {journal} {Physical review letters}\ }\textbf {\bibinfo {volume} {104}},\ \bibinfo {pages} {195001} (\bibinfo {year} {2010})}\BibitemShut {NoStop}%
\bibitem [{\citenamefont {Zhdanov}\ \emph {et~al.}(2009)\citenamefont {Zhdanov}, \citenamefont {Ivlev},\ and\ \citenamefont {Morfill}}]{zhdanov2009mode}%
  \BibitemOpen
  \bibfield  {author} {\bibinfo {author} {\bibfnamefont {S.}~\bibnamefont {Zhdanov}}, \bibinfo {author} {\bibfnamefont {A.}~\bibnamefont {Ivlev}}, \ and\ \bibinfo {author} {\bibfnamefont {G.}~\bibnamefont {Morfill}},\ }\href@noop {} {\bibfield  {journal} {\bibinfo  {journal} {Physics of Plasmas}\ }\textbf {\bibinfo {volume} {16}},\ \bibinfo {pages} {083706} (\bibinfo {year} {2009})}\BibitemShut {NoStop}%
\bibitem [{\citenamefont {Konopka}\ \emph {et~al.}(2000)\citenamefont {Konopka}, \citenamefont {Morfill},\ and\ \citenamefont {Ratke}}]{konopka2000measurement}%
  \BibitemOpen
  \bibfield  {author} {\bibinfo {author} {\bibfnamefont {U.}~\bibnamefont {Konopka}}, \bibinfo {author} {\bibfnamefont {G.}~\bibnamefont {Morfill}}, \ and\ \bibinfo {author} {\bibfnamefont {L.}~\bibnamefont {Ratke}},\ }\href@noop {} {\bibfield  {journal} {\bibinfo  {journal} {Physical review letters}\ }\textbf {\bibinfo {volume} {84}},\ \bibinfo {pages} {891} (\bibinfo {year} {2000})}\BibitemShut {NoStop}%
\bibitem [{\citenamefont {Gogia}\ and\ \citenamefont {Burton}(2017)}]{gogia2017emergent}%
  \BibitemOpen
  \bibfield  {author} {\bibinfo {author} {\bibfnamefont {G.}~\bibnamefont {Gogia}}\ and\ \bibinfo {author} {\bibfnamefont {J.~C.}\ \bibnamefont {Burton}},\ }\href@noop {} {\bibfield  {journal} {\bibinfo  {journal} {Physical Review Letters}\ }\textbf {\bibinfo {volume} {119}},\ \bibinfo {pages} {178004} (\bibinfo {year} {2017})}\BibitemShut {NoStop}%
\bibitem [{\citenamefont {Greiner}\ \emph {et~al.}(2018)\citenamefont {Greiner}, \citenamefont {Melzer}, \citenamefont {Tadsen}, \citenamefont {Groth}, \citenamefont {Killer}, \citenamefont {Kirchschlager}, \citenamefont {Wieben}, \citenamefont {Pilch}, \citenamefont {Kr{\"u}ger}, \citenamefont {Block} \emph {et~al.}}]{greiner2018diagnostics}%
  \BibitemOpen
  \bibfield  {author} {\bibinfo {author} {\bibfnamefont {F.}~\bibnamefont {Greiner}}, \bibinfo {author} {\bibfnamefont {A.}~\bibnamefont {Melzer}}, \bibinfo {author} {\bibfnamefont {B.}~\bibnamefont {Tadsen}}, \bibinfo {author} {\bibfnamefont {S.}~\bibnamefont {Groth}}, \bibinfo {author} {\bibfnamefont {C.}~\bibnamefont {Killer}}, \bibinfo {author} {\bibfnamefont {F.}~\bibnamefont {Kirchschlager}}, \bibinfo {author} {\bibfnamefont {F.}~\bibnamefont {Wieben}}, \bibinfo {author} {\bibfnamefont {I.}~\bibnamefont {Pilch}}, \bibinfo {author} {\bibfnamefont {H.}~\bibnamefont {Kr{\"u}ger}}, \bibinfo {author} {\bibfnamefont {D.}~\bibnamefont {Block}},  \emph {et~al.},\ }\href@noop {} {\bibfield  {journal} {\bibinfo  {journal} {The European Physical Journal D}\ }\textbf {\bibinfo {volume} {72}},\ \bibinfo {pages} {1} (\bibinfo {year} {2018})}\BibitemShut {NoStop}%
\bibitem [{\citenamefont {Lampe}\ \emph {et~al.}(2000)\citenamefont {Lampe}, \citenamefont {Joyce}, \citenamefont {Ganguli},\ and\ \citenamefont {Gavrishchaka}}]{lampe2000interactions}%
  \BibitemOpen
  \bibfield  {author} {\bibinfo {author} {\bibfnamefont {M.}~\bibnamefont {Lampe}}, \bibinfo {author} {\bibfnamefont {G.}~\bibnamefont {Joyce}}, \bibinfo {author} {\bibfnamefont {G.}~\bibnamefont {Ganguli}}, \ and\ \bibinfo {author} {\bibfnamefont {V.}~\bibnamefont {Gavrishchaka}},\ }\href@noop {} {\bibfield  {journal} {\bibinfo  {journal} {Physics of Plasmas}\ }\textbf {\bibinfo {volume} {7}},\ \bibinfo {pages} {3851} (\bibinfo {year} {2000})}\BibitemShut {NoStop}%
\bibitem [{\citenamefont {Gopalakrishnan}\ and\ \citenamefont {Hogan~Jr}(2012)}]{gopalakrishnan2012coulomb}%
  \BibitemOpen
  \bibfield  {author} {\bibinfo {author} {\bibfnamefont {R.}~\bibnamefont {Gopalakrishnan}}\ and\ \bibinfo {author} {\bibfnamefont {C.~J.}\ \bibnamefont {Hogan~Jr}},\ }\href@noop {} {\bibfield  {journal} {\bibinfo  {journal} {Physical Review E}\ }\textbf {\bibinfo {volume} {85}},\ \bibinfo {pages} {026410} (\bibinfo {year} {2012})}\BibitemShut {NoStop}%
\bibitem [{\citenamefont {Ignatov}(1997)}]{ignatov1997interaction}%
  \BibitemOpen
  \bibfield  {author} {\bibinfo {author} {\bibfnamefont {A.}~\bibnamefont {Ignatov}},\ }\href@noop {} {\bibfield  {journal} {\bibinfo  {journal} {Le Journal de Physique IV}\ }\textbf {\bibinfo {volume} {7}},\ \bibinfo {pages} {C4} (\bibinfo {year} {1997})}\BibitemShut {NoStop}%
\bibitem [{\citenamefont {Fulton}\ \emph {et~al.}(2018)\citenamefont {Fulton}, \citenamefont {Petigura}, \citenamefont {Blunt},\ and\ \citenamefont {Sinukoff}}]{fulton2018radvel}%
  \BibitemOpen
  \bibfield  {author} {\bibinfo {author} {\bibfnamefont {B.~J.}\ \bibnamefont {Fulton}}, \bibinfo {author} {\bibfnamefont {E.~A.}\ \bibnamefont {Petigura}}, \bibinfo {author} {\bibfnamefont {S.}~\bibnamefont {Blunt}}, \ and\ \bibinfo {author} {\bibfnamefont {E.}~\bibnamefont {Sinukoff}},\ }\href@noop {} {\bibfield  {journal} {\bibinfo  {journal} {Publications of the Astronomical Society of the Pacific}\ }\textbf {\bibinfo {volume} {130}},\ \bibinfo {pages} {044504} (\bibinfo {year} {2018})}\BibitemShut {NoStop}%
\bibitem [{\citenamefont {Mayor}\ and\ \citenamefont {Queloz}(1995)}]{mayor1995jupiter}%
  \BibitemOpen
  \bibfield  {author} {\bibinfo {author} {\bibfnamefont {M.}~\bibnamefont {Mayor}}\ and\ \bibinfo {author} {\bibfnamefont {D.}~\bibnamefont {Queloz}},\ }\href@noop {} {\bibfield  {journal} {\bibinfo  {journal} {nature}\ }\textbf {\bibinfo {volume} {378}},\ \bibinfo {pages} {355} (\bibinfo {year} {1995})}\BibitemShut {NoStop}%
\bibitem [{\citenamefont {Bapst}\ \emph {et~al.}(2020)\citenamefont {Bapst}, \citenamefont {Keck}, \citenamefont {Grabska-Barwi{\'n}ska}, \citenamefont {Donner}, \citenamefont {Cubuk}, \citenamefont {Schoenholz}, \citenamefont {Obika}, \citenamefont {Nelson}, \citenamefont {Back}, \citenamefont {Hassabis} \emph {et~al.}}]{bapst2020unveiling}%
  \BibitemOpen
  \bibfield  {author} {\bibinfo {author} {\bibfnamefont {V.}~\bibnamefont {Bapst}}, \bibinfo {author} {\bibfnamefont {T.}~\bibnamefont {Keck}}, \bibinfo {author} {\bibfnamefont {A.}~\bibnamefont {Grabska-Barwi{\'n}ska}}, \bibinfo {author} {\bibfnamefont {C.}~\bibnamefont {Donner}}, \bibinfo {author} {\bibfnamefont {E.~D.}\ \bibnamefont {Cubuk}}, \bibinfo {author} {\bibfnamefont {S.~S.}\ \bibnamefont {Schoenholz}}, \bibinfo {author} {\bibfnamefont {A.}~\bibnamefont {Obika}}, \bibinfo {author} {\bibfnamefont {A.~W.}\ \bibnamefont {Nelson}}, \bibinfo {author} {\bibfnamefont {T.}~\bibnamefont {Back}}, \bibinfo {author} {\bibfnamefont {D.}~\bibnamefont {Hassabis}},  \emph {et~al.},\ }\href@noop {} {\bibfield  {journal} {\bibinfo  {journal} {Nature Physics}\ }\textbf {\bibinfo {volume} {16}},\ \bibinfo {pages} {448} (\bibinfo {year} {2020})}\BibitemShut {NoStop}%
\bibitem [{\citenamefont {Colen}\ \emph {et~al.}(2021)\citenamefont {Colen}, \citenamefont {Han}, \citenamefont {Zhang}, \citenamefont {Redford}, \citenamefont {Lemma}, \citenamefont {Morgan}, \citenamefont {Ruijgrok}, \citenamefont {Adkins}, \citenamefont {Bryant}, \citenamefont {Dogic} \emph {et~al.}}]{colen2021machine}%
  \BibitemOpen
  \bibfield  {author} {\bibinfo {author} {\bibfnamefont {J.}~\bibnamefont {Colen}}, \bibinfo {author} {\bibfnamefont {M.}~\bibnamefont {Han}}, \bibinfo {author} {\bibfnamefont {R.}~\bibnamefont {Zhang}}, \bibinfo {author} {\bibfnamefont {S.~A.}\ \bibnamefont {Redford}}, \bibinfo {author} {\bibfnamefont {L.~M.}\ \bibnamefont {Lemma}}, \bibinfo {author} {\bibfnamefont {L.}~\bibnamefont {Morgan}}, \bibinfo {author} {\bibfnamefont {P.~V.}\ \bibnamefont {Ruijgrok}}, \bibinfo {author} {\bibfnamefont {R.}~\bibnamefont {Adkins}}, \bibinfo {author} {\bibfnamefont {Z.}~\bibnamefont {Bryant}}, \bibinfo {author} {\bibfnamefont {Z.}~\bibnamefont {Dogic}},  \emph {et~al.},\ }\href@noop {} {\bibfield  {journal} {\bibinfo  {journal} {Proceedings of the National Academy of Sciences}\ }\textbf {\bibinfo {volume} {118}},\ \bibinfo {pages} {e2016708118} (\bibinfo {year} {2021})}\BibitemShut {NoStop}%
\bibitem [{\citenamefont {Tah}\ \emph {et~al.}(2022)\citenamefont {Tah}, \citenamefont {Ridout},\ and\ \citenamefont {Liu}}]{tah2022fragility}%
  \BibitemOpen
  \bibfield  {author} {\bibinfo {author} {\bibfnamefont {I.}~\bibnamefont {Tah}}, \bibinfo {author} {\bibfnamefont {S.~A.}\ \bibnamefont {Ridout}}, \ and\ \bibinfo {author} {\bibfnamefont {A.~J.}\ \bibnamefont {Liu}},\ }\href@noop {} {\bibfield  {journal} {\bibinfo  {journal} {The Journal of Chemical Physics}\ }\textbf {\bibinfo {volume} {157}},\ \bibinfo {pages} {124501} (\bibinfo {year} {2022})}\BibitemShut {NoStop}%
\bibitem [{\citenamefont {Daniels}\ \emph {et~al.}(2019)\citenamefont {Daniels}, \citenamefont {Ryu},\ and\ \citenamefont {Nemenman}}]{daniels2019automated}%
  \BibitemOpen
  \bibfield  {author} {\bibinfo {author} {\bibfnamefont {B.~C.}\ \bibnamefont {Daniels}}, \bibinfo {author} {\bibfnamefont {W.~S.}\ \bibnamefont {Ryu}}, \ and\ \bibinfo {author} {\bibfnamefont {I.}~\bibnamefont {Nemenman}},\ }\href@noop {} {\bibfield  {journal} {\bibinfo  {journal} {Proceedings of the National Academy of Sciences}\ }\textbf {\bibinfo {volume} {116}},\ \bibinfo {pages} {7226} (\bibinfo {year} {2019})}\BibitemShut {NoStop}%
\bibitem [{\citenamefont {Anstine}\ and\ \citenamefont {Isayev}(2023)}]{anstine2023machine}%
  \BibitemOpen
  \bibfield  {author} {\bibinfo {author} {\bibfnamefont {D.~M.}\ \bibnamefont {Anstine}}\ and\ \bibinfo {author} {\bibfnamefont {O.}~\bibnamefont {Isayev}},\ }\href@noop {} {\bibfield  {journal} {\bibinfo  {journal} {The Journal of Physical Chemistry A}\ }\textbf {\bibinfo {volume} {127}},\ \bibinfo {pages} {2417} (\bibinfo {year} {2023})}\BibitemShut {NoStop}%
\bibitem [{\citenamefont {Champion}\ \emph {et~al.}(2019)\citenamefont {Champion}, \citenamefont {Lusch}, \citenamefont {Kutz},\ and\ \citenamefont {Brunton}}]{champion2019data}%
  \BibitemOpen
  \bibfield  {author} {\bibinfo {author} {\bibfnamefont {K.}~\bibnamefont {Champion}}, \bibinfo {author} {\bibfnamefont {B.}~\bibnamefont {Lusch}}, \bibinfo {author} {\bibfnamefont {J.~N.}\ \bibnamefont {Kutz}}, \ and\ \bibinfo {author} {\bibfnamefont {S.~L.}\ \bibnamefont {Brunton}},\ }\href@noop {} {\bibfield  {journal} {\bibinfo  {journal} {Proceedings of the National Academy of Sciences}\ }\textbf {\bibinfo {volume} {116}},\ \bibinfo {pages} {22445} (\bibinfo {year} {2019})}\BibitemShut {NoStop}%
\bibitem [{\citenamefont {Brunton}\ \emph {et~al.}(2016)\citenamefont {Brunton}, \citenamefont {Proctor},\ and\ \citenamefont {Kutz}}]{brunton2016discovering}%
  \BibitemOpen
  \bibfield  {author} {\bibinfo {author} {\bibfnamefont {S.~L.}\ \bibnamefont {Brunton}}, \bibinfo {author} {\bibfnamefont {J.~L.}\ \bibnamefont {Proctor}}, \ and\ \bibinfo {author} {\bibfnamefont {J.~N.}\ \bibnamefont {Kutz}},\ }\href@noop {} {\bibfield  {journal} {\bibinfo  {journal} {Proceedings of the national academy of sciences}\ }\textbf {\bibinfo {volume} {113}},\ \bibinfo {pages} {3932} (\bibinfo {year} {2016})}\BibitemShut {NoStop}%
\bibitem [{\citenamefont {Rudy}\ \emph {et~al.}(2017)\citenamefont {Rudy}, \citenamefont {Brunton}, \citenamefont {Proctor},\ and\ \citenamefont {Kutz}}]{rudy2017data}%
  \BibitemOpen
  \bibfield  {author} {\bibinfo {author} {\bibfnamefont {S.~H.}\ \bibnamefont {Rudy}}, \bibinfo {author} {\bibfnamefont {S.~L.}\ \bibnamefont {Brunton}}, \bibinfo {author} {\bibfnamefont {J.~L.}\ \bibnamefont {Proctor}}, \ and\ \bibinfo {author} {\bibfnamefont {J.~N.}\ \bibnamefont {Kutz}},\ }\href@noop {} {\bibfield  {journal} {\bibinfo  {journal} {Science advances}\ }\textbf {\bibinfo {volume} {3}},\ \bibinfo {pages} {e1602614} (\bibinfo {year} {2017})}\BibitemShut {NoStop}%
\bibitem [{\citenamefont {Br{\"u}ckner}\ \emph {et~al.}(2020)\citenamefont {Br{\"u}ckner}, \citenamefont {Ronceray},\ and\ \citenamefont {Broedersz}}]{bruckner2020inferring}%
  \BibitemOpen
  \bibfield  {author} {\bibinfo {author} {\bibfnamefont {D.~B.}\ \bibnamefont {Br{\"u}ckner}}, \bibinfo {author} {\bibfnamefont {P.}~\bibnamefont {Ronceray}}, \ and\ \bibinfo {author} {\bibfnamefont {C.~P.}\ \bibnamefont {Broedersz}},\ }\href@noop {} {\bibfield  {journal} {\bibinfo  {journal} {Physical review letters}\ }\textbf {\bibinfo {volume} {125}},\ \bibinfo {pages} {058103} (\bibinfo {year} {2020})}\BibitemShut {NoStop}%
\bibitem [{\citenamefont {Lemos}\ \emph {et~al.}(2023)\citenamefont {Lemos}, \citenamefont {Jeffrey}, \citenamefont {Cranmer}, \citenamefont {Ho},\ and\ \citenamefont {Battaglia}}]{lemos2023rediscovering}%
  \BibitemOpen
  \bibfield  {author} {\bibinfo {author} {\bibfnamefont {P.}~\bibnamefont {Lemos}}, \bibinfo {author} {\bibfnamefont {N.}~\bibnamefont {Jeffrey}}, \bibinfo {author} {\bibfnamefont {M.}~\bibnamefont {Cranmer}}, \bibinfo {author} {\bibfnamefont {S.}~\bibnamefont {Ho}}, \ and\ \bibinfo {author} {\bibfnamefont {P.}~\bibnamefont {Battaglia}},\ }\href@noop {} {\bibfield  {journal} {\bibinfo  {journal} {Machine Learning: Science and Technology}\ }\textbf {\bibinfo {volume} {4}},\ \bibinfo {pages} {045002} (\bibinfo {year} {2023})}\BibitemShut {NoStop}%
\bibitem [{\citenamefont {Pandarinath}\ \emph {et~al.}(2018)\citenamefont {Pandarinath}, \citenamefont {O’Shea}, \citenamefont {Collins}, \citenamefont {Jozefowicz}, \citenamefont {Stavisky}, \citenamefont {Kao}, \citenamefont {Trautmann}, \citenamefont {Kaufman}, \citenamefont {Ryu}, \citenamefont {Hochberg} \emph {et~al.}}]{pandarinath2018inferring}%
  \BibitemOpen
  \bibfield  {author} {\bibinfo {author} {\bibfnamefont {C.}~\bibnamefont {Pandarinath}}, \bibinfo {author} {\bibfnamefont {D.~J.}\ \bibnamefont {O’Shea}}, \bibinfo {author} {\bibfnamefont {J.}~\bibnamefont {Collins}}, \bibinfo {author} {\bibfnamefont {R.}~\bibnamefont {Jozefowicz}}, \bibinfo {author} {\bibfnamefont {S.~D.}\ \bibnamefont {Stavisky}}, \bibinfo {author} {\bibfnamefont {J.~C.}\ \bibnamefont {Kao}}, \bibinfo {author} {\bibfnamefont {E.~M.}\ \bibnamefont {Trautmann}}, \bibinfo {author} {\bibfnamefont {M.~T.}\ \bibnamefont {Kaufman}}, \bibinfo {author} {\bibfnamefont {S.~I.}\ \bibnamefont {Ryu}}, \bibinfo {author} {\bibfnamefont {L.~R.}\ \bibnamefont {Hochberg}},  \emph {et~al.},\ }\href@noop {} {\bibfield  {journal} {\bibinfo  {journal} {Nature methods}\ }\textbf {\bibinfo {volume} {15}},\ \bibinfo {pages} {805} (\bibinfo {year} {2018})}\BibitemShut {NoStop}%
\bibitem [{\citenamefont {Daniels}\ and\ \citenamefont {Nemenman}(2015)}]{daniels2015automated}%
  \BibitemOpen
  \bibfield  {author} {\bibinfo {author} {\bibfnamefont {B.~C.}\ \bibnamefont {Daniels}}\ and\ \bibinfo {author} {\bibfnamefont {I.}~\bibnamefont {Nemenman}},\ }\href@noop {} {\bibfield  {journal} {\bibinfo  {journal} {Nature communications}\ }\textbf {\bibinfo {volume} {6}},\ \bibinfo {pages} {8133} (\bibinfo {year} {2015})}\BibitemShut {NoStop}%
\bibitem [{\citenamefont {Chen}\ \emph {et~al.}(2022)\citenamefont {Chen}, \citenamefont {Huang}, \citenamefont {Raghupathi}, \citenamefont {Chandratreya}, \citenamefont {Du},\ and\ \citenamefont {Lipson}}]{chen2022automated}%
  \BibitemOpen
  \bibfield  {author} {\bibinfo {author} {\bibfnamefont {B.}~\bibnamefont {Chen}}, \bibinfo {author} {\bibfnamefont {K.}~\bibnamefont {Huang}}, \bibinfo {author} {\bibfnamefont {S.}~\bibnamefont {Raghupathi}}, \bibinfo {author} {\bibfnamefont {I.}~\bibnamefont {Chandratreya}}, \bibinfo {author} {\bibfnamefont {Q.}~\bibnamefont {Du}}, \ and\ \bibinfo {author} {\bibfnamefont {H.}~\bibnamefont {Lipson}},\ }\href@noop {} {\bibfield  {journal} {\bibinfo  {journal} {Nature Computational Science}\ }\textbf {\bibinfo {volume} {2}},\ \bibinfo {pages} {433} (\bibinfo {year} {2022})}\BibitemShut {NoStop}%
\bibitem [{\citenamefont {Supekar}\ \emph {et~al.}(2023)\citenamefont {Supekar}, \citenamefont {Song}, \citenamefont {Hastewell}, \citenamefont {Choi}, \citenamefont {Mietke},\ and\ \citenamefont {Dunkel}}]{supekar2023learning}%
  \BibitemOpen
  \bibfield  {author} {\bibinfo {author} {\bibfnamefont {R.}~\bibnamefont {Supekar}}, \bibinfo {author} {\bibfnamefont {B.}~\bibnamefont {Song}}, \bibinfo {author} {\bibfnamefont {A.}~\bibnamefont {Hastewell}}, \bibinfo {author} {\bibfnamefont {G.~P.}\ \bibnamefont {Choi}}, \bibinfo {author} {\bibfnamefont {A.}~\bibnamefont {Mietke}}, \ and\ \bibinfo {author} {\bibfnamefont {J.}~\bibnamefont {Dunkel}},\ }\href@noop {} {\bibfield  {journal} {\bibinfo  {journal} {Proceedings of the National Academy of Sciences}\ }\textbf {\bibinfo {volume} {120}},\ \bibinfo {pages} {e2206994120} (\bibinfo {year} {2023})}\BibitemShut {NoStop}%
\bibitem [{\citenamefont {Raissi}\ \emph {et~al.}(2020)\citenamefont {Raissi}, \citenamefont {Yazdani},\ and\ \citenamefont {Karniadakis}}]{raissi2020hidden}%
  \BibitemOpen
  \bibfield  {author} {\bibinfo {author} {\bibfnamefont {M.}~\bibnamefont {Raissi}}, \bibinfo {author} {\bibfnamefont {A.}~\bibnamefont {Yazdani}}, \ and\ \bibinfo {author} {\bibfnamefont {G.~E.}\ \bibnamefont {Karniadakis}},\ }\href@noop {} {\bibfield  {journal} {\bibinfo  {journal} {Science}\ }\textbf {\bibinfo {volume} {367}},\ \bibinfo {pages} {1026} (\bibinfo {year} {2020})}\BibitemShut {NoStop}%
\bibitem [{\citenamefont {Ruiz-Garcia}\ \emph {et~al.}(2024)\citenamefont {Ruiz-Garcia}, \citenamefont {Barriuso~G}, \citenamefont {Alexander}, \citenamefont {Aarts}, \citenamefont {Ghiringhelli},\ and\ \citenamefont {Valeriani}}]{ruiz2024discovering}%
  \BibitemOpen
  \bibfield  {author} {\bibinfo {author} {\bibfnamefont {M.}~\bibnamefont {Ruiz-Garcia}}, \bibinfo {author} {\bibfnamefont {C.~M.}\ \bibnamefont {Barriuso~G}}, \bibinfo {author} {\bibfnamefont {L.~C.}\ \bibnamefont {Alexander}}, \bibinfo {author} {\bibfnamefont {D.~G.}\ \bibnamefont {Aarts}}, \bibinfo {author} {\bibfnamefont {L.~M.}\ \bibnamefont {Ghiringhelli}}, \ and\ \bibinfo {author} {\bibfnamefont {C.}~\bibnamefont {Valeriani}},\ }\href@noop {} {\bibfield  {journal} {\bibinfo  {journal} {Physical Review E}\ }\textbf {\bibinfo {volume} {109}},\ \bibinfo {pages} {064611} (\bibinfo {year} {2024})}\BibitemShut {NoStop}%
\bibitem [{\citenamefont {Battiston}\ \emph {et~al.}(2021)\citenamefont {Battiston}, \citenamefont {Amico}, \citenamefont {Barrat}, \citenamefont {Bianconi}, \citenamefont {Ferraz~de Arruda}, \citenamefont {Franceschiello}, \citenamefont {Iacopini}, \citenamefont {K{\'e}fi}, \citenamefont {Latora}, \citenamefont {Moreno} \emph {et~al.}}]{battiston2021physics}%
  \BibitemOpen
  \bibfield  {author} {\bibinfo {author} {\bibfnamefont {F.}~\bibnamefont {Battiston}}, \bibinfo {author} {\bibfnamefont {E.}~\bibnamefont {Amico}}, \bibinfo {author} {\bibfnamefont {A.}~\bibnamefont {Barrat}}, \bibinfo {author} {\bibfnamefont {G.}~\bibnamefont {Bianconi}}, \bibinfo {author} {\bibfnamefont {G.}~\bibnamefont {Ferraz~de Arruda}}, \bibinfo {author} {\bibfnamefont {B.}~\bibnamefont {Franceschiello}}, \bibinfo {author} {\bibfnamefont {I.}~\bibnamefont {Iacopini}}, \bibinfo {author} {\bibfnamefont {S.}~\bibnamefont {K{\'e}fi}}, \bibinfo {author} {\bibfnamefont {V.}~\bibnamefont {Latora}}, \bibinfo {author} {\bibfnamefont {Y.}~\bibnamefont {Moreno}},  \emph {et~al.},\ }\href@noop {} {\bibfield  {journal} {\bibinfo  {journal} {Nature Physics}\ }\textbf {\bibinfo {volume} {17}},\ \bibinfo {pages} {1093} (\bibinfo {year} {2021})}\BibitemShut {NoStop}%
\bibitem [{\citenamefont {Karniadakis}\ \emph {et~al.}(2021)\citenamefont {Karniadakis}, \citenamefont {Kevrekidis}, \citenamefont {Lu}, \citenamefont {Perdikaris}, \citenamefont {Wang},\ and\ \citenamefont {Yang}}]{karniadakis2021physics}%
  \BibitemOpen
  \bibfield  {author} {\bibinfo {author} {\bibfnamefont {G.~E.}\ \bibnamefont {Karniadakis}}, \bibinfo {author} {\bibfnamefont {I.~G.}\ \bibnamefont {Kevrekidis}}, \bibinfo {author} {\bibfnamefont {L.}~\bibnamefont {Lu}}, \bibinfo {author} {\bibfnamefont {P.}~\bibnamefont {Perdikaris}}, \bibinfo {author} {\bibfnamefont {S.}~\bibnamefont {Wang}}, \ and\ \bibinfo {author} {\bibfnamefont {L.}~\bibnamefont {Yang}},\ }\href@noop {} {\bibfield  {journal} {\bibinfo  {journal} {Nature Reviews Physics}\ }\textbf {\bibinfo {volume} {3}},\ \bibinfo {pages} {422} (\bibinfo {year} {2021})}\BibitemShut {NoStop}%
\bibitem [{\citenamefont {Cichos}\ \emph {et~al.}(2020)\citenamefont {Cichos}, \citenamefont {Gustavsson}, \citenamefont {Mehlig},\ and\ \citenamefont {Volpe}}]{cichos2020machine}%
  \BibitemOpen
  \bibfield  {author} {\bibinfo {author} {\bibfnamefont {F.}~\bibnamefont {Cichos}}, \bibinfo {author} {\bibfnamefont {K.}~\bibnamefont {Gustavsson}}, \bibinfo {author} {\bibfnamefont {B.}~\bibnamefont {Mehlig}}, \ and\ \bibinfo {author} {\bibfnamefont {G.}~\bibnamefont {Volpe}},\ }\href@noop {} {\bibfield  {journal} {\bibinfo  {journal} {Nature Machine Intelligence}\ }\textbf {\bibinfo {volume} {2}},\ \bibinfo {pages} {94} (\bibinfo {year} {2020})}\BibitemShut {NoStop}%
\bibitem [{\citenamefont {Falk}\ \emph {et~al.}(2021)\citenamefont {Falk}, \citenamefont {Alizadehyazdi}, \citenamefont {Jaeger},\ and\ \citenamefont {Murugan}}]{falk2021learning}%
  \BibitemOpen
  \bibfield  {author} {\bibinfo {author} {\bibfnamefont {M.~J.}\ \bibnamefont {Falk}}, \bibinfo {author} {\bibfnamefont {V.}~\bibnamefont {Alizadehyazdi}}, \bibinfo {author} {\bibfnamefont {H.}~\bibnamefont {Jaeger}}, \ and\ \bibinfo {author} {\bibfnamefont {A.}~\bibnamefont {Murugan}},\ }\href@noop {} {\bibfield  {journal} {\bibinfo  {journal} {Physical Review Research}\ }\textbf {\bibinfo {volume} {3}},\ \bibinfo {pages} {033291} (\bibinfo {year} {2021})}\BibitemShut {NoStop}%
\bibitem [{\citenamefont {Han}\ \emph {et~al.}(2022)\citenamefont {Han}, \citenamefont {Kammer},\ and\ \citenamefont {Fink}}]{han2022learning}%
  \BibitemOpen
  \bibfield  {author} {\bibinfo {author} {\bibfnamefont {Z.}~\bibnamefont {Han}}, \bibinfo {author} {\bibfnamefont {D.~S.}\ \bibnamefont {Kammer}}, \ and\ \bibinfo {author} {\bibfnamefont {O.}~\bibnamefont {Fink}},\ }\href@noop {} {\bibfield  {journal} {\bibinfo  {journal} {PNAS nexus}\ }\textbf {\bibinfo {volume} {1}},\ \bibinfo {pages} {pgac264} (\bibinfo {year} {2022})}\BibitemShut {NoStop}%
\bibitem [{\citenamefont {Ishihara}\ and\ \citenamefont {Vladimirov}(1997)}]{ishihara1997wake}%
  \BibitemOpen
  \bibfield  {author} {\bibinfo {author} {\bibfnamefont {O.}~\bibnamefont {Ishihara}}\ and\ \bibinfo {author} {\bibfnamefont {S.~V.}\ \bibnamefont {Vladimirov}},\ }\href@noop {} {\bibfield  {journal} {\bibinfo  {journal} {Physics of Plasmas}\ }\textbf {\bibinfo {volume} {4}},\ \bibinfo {pages} {69} (\bibinfo {year} {1997})}\BibitemShut {NoStop}%
\bibitem [{\citenamefont {Ludwig}\ \emph {et~al.}(2012)\citenamefont {Ludwig}, \citenamefont {Miloch}, \citenamefont {K{\"a}hlert},\ and\ \citenamefont {Bonitz}}]{ludwig2012wake}%
  \BibitemOpen
  \bibfield  {author} {\bibinfo {author} {\bibfnamefont {P.}~\bibnamefont {Ludwig}}, \bibinfo {author} {\bibfnamefont {W.~J.}\ \bibnamefont {Miloch}}, \bibinfo {author} {\bibfnamefont {H.}~\bibnamefont {K{\"a}hlert}}, \ and\ \bibinfo {author} {\bibfnamefont {M.}~\bibnamefont {Bonitz}},\ }\href@noop {} {\bibfield  {journal} {\bibinfo  {journal} {New Journal of Physics}\ }\textbf {\bibinfo {volume} {14}},\ \bibinfo {pages} {053016} (\bibinfo {year} {2012})}\BibitemShut {NoStop}%
\bibitem [{\citenamefont {Yu}\ and\ \citenamefont {Burton}(2023)}]{yu20233d}%
  \BibitemOpen
  \bibfield  {author} {\bibinfo {author} {\bibfnamefont {W.}~\bibnamefont {Yu}}\ and\ \bibinfo {author} {\bibfnamefont {J.~C.}\ \bibnamefont {Burton}},\ }\href@noop {} {\bibfield  {journal} {\bibinfo  {journal} {Physics of Plasmas}\ }\textbf {\bibinfo {volume} {30}},\ \bibinfo {pages} {063701} (\bibinfo {year} {2023})}\BibitemShut {NoStop}%
\bibitem [{\citenamefont {Goree}(1994)}]{goree1994charging}%
  \BibitemOpen
  \bibfield  {author} {\bibinfo {author} {\bibfnamefont {J.}~\bibnamefont {Goree}},\ }\href@noop {} {\bibfield  {journal} {\bibinfo  {journal} {Plasma Sources Science and Technology}\ }\textbf {\bibinfo {volume} {3}},\ \bibinfo {pages} {400} (\bibinfo {year} {1994})}\BibitemShut {NoStop}%
\bibitem [{\citenamefont {Harper}\ \emph {et~al.}(2020)\citenamefont {Harper}, \citenamefont {Gogia}, \citenamefont {Wu}, \citenamefont {Laseter},\ and\ \citenamefont {Burton}}]{harper2020origin}%
  \BibitemOpen
  \bibfield  {author} {\bibinfo {author} {\bibfnamefont {J.~M.}\ \bibnamefont {Harper}}, \bibinfo {author} {\bibfnamefont {G.}~\bibnamefont {Gogia}}, \bibinfo {author} {\bibfnamefont {B.}~\bibnamefont {Wu}}, \bibinfo {author} {\bibfnamefont {Z.}~\bibnamefont {Laseter}}, \ and\ \bibinfo {author} {\bibfnamefont {J.~C.}\ \bibnamefont {Burton}},\ }\href@noop {} {\bibfield  {journal} {\bibinfo  {journal} {Physical Review Research}\ }\textbf {\bibinfo {volume} {2}},\ \bibinfo {pages} {033500} (\bibinfo {year} {2020})}\BibitemShut {NoStop}%
\bibitem [{\citenamefont {Nitter}(1996)}]{nitter1996levitation}%
  \BibitemOpen
  \bibfield  {author} {\bibinfo {author} {\bibfnamefont {T.}~\bibnamefont {Nitter}},\ }\href@noop {} {\bibfield  {journal} {\bibinfo  {journal} {Plasma Sources Science and Technology}\ }\textbf {\bibinfo {volume} {5}},\ \bibinfo {pages} {93} (\bibinfo {year} {1996})}\BibitemShut {NoStop}%
\bibitem [{\citenamefont {Epstein}(1924)}]{epstein1924resistance}%
  \BibitemOpen
  \bibfield  {author} {\bibinfo {author} {\bibfnamefont {P.~S.}\ \bibnamefont {Epstein}},\ }\href@noop {} {\bibfield  {journal} {\bibinfo  {journal} {Physical Review}\ }\textbf {\bibinfo {volume} {23}},\ \bibinfo {pages} {710} (\bibinfo {year} {1924})}\BibitemShut {NoStop}%
\bibitem [{\citenamefont {Gurevich}\ \emph {et~al.}(2019)\citenamefont {Gurevich}, \citenamefont {Reinbold},\ and\ \citenamefont {Grigoriev}}]{gurevich2019robust}%
  \BibitemOpen
  \bibfield  {author} {\bibinfo {author} {\bibfnamefont {D.~R.}\ \bibnamefont {Gurevich}}, \bibinfo {author} {\bibfnamefont {P.~A.}\ \bibnamefont {Reinbold}}, \ and\ \bibinfo {author} {\bibfnamefont {R.~O.}\ \bibnamefont {Grigoriev}},\ }\href@noop {} {\bibfield  {journal} {\bibinfo  {journal} {Chaos: An Interdisciplinary Journal of Nonlinear Science}\ }\textbf {\bibinfo {volume} {29}},\ \bibinfo {pages} {103113} (\bibinfo {year} {2019})}\BibitemShut {NoStop}%
\bibitem [{\citenamefont {Vladimirov}\ and\ \citenamefont {Nambu}(1995)}]{vladimirov1995attraction}%
  \BibitemOpen
  \bibfield  {author} {\bibinfo {author} {\bibfnamefont {S.~V.}\ \bibnamefont {Vladimirov}}\ and\ \bibinfo {author} {\bibfnamefont {M.}~\bibnamefont {Nambu}},\ }\href@noop {} {\bibfield  {journal} {\bibinfo  {journal} {Physical Review E}\ }\textbf {\bibinfo {volume} {52}},\ \bibinfo {pages} {R2172} (\bibinfo {year} {1995})}\BibitemShut {NoStop}%
\bibitem [{\citenamefont {Kryuchkov}\ \emph {et~al.}(2020)\citenamefont {Kryuchkov}, \citenamefont {Mistryukova}, \citenamefont {Sapelkin},\ and\ \citenamefont {Yurchenko}}]{kryuchkov2020strange}%
  \BibitemOpen
  \bibfield  {author} {\bibinfo {author} {\bibfnamefont {N.~P.}\ \bibnamefont {Kryuchkov}}, \bibinfo {author} {\bibfnamefont {L.~A.}\ \bibnamefont {Mistryukova}}, \bibinfo {author} {\bibfnamefont {A.~V.}\ \bibnamefont {Sapelkin}}, \ and\ \bibinfo {author} {\bibfnamefont {S.~O.}\ \bibnamefont {Yurchenko}},\ }\href@noop {} {\bibfield  {journal} {\bibinfo  {journal} {Physical Review E}\ }\textbf {\bibinfo {volume} {101}},\ \bibinfo {pages} {063205} (\bibinfo {year} {2020})}\BibitemShut {NoStop}%
\bibitem [{\citenamefont {Bohm}\ and\ \citenamefont {Gross}(1950)}]{bohm1950effects}%
  \BibitemOpen
  \bibfield  {author} {\bibinfo {author} {\bibfnamefont {D.}~\bibnamefont {Bohm}}\ and\ \bibinfo {author} {\bibfnamefont {E.~P.}\ \bibnamefont {Gross}},\ }\href@noop {} {\bibfield  {journal} {\bibinfo  {journal} {Physical Review}\ }\textbf {\bibinfo {volume} {79}},\ \bibinfo {pages} {992} (\bibinfo {year} {1950})}\BibitemShut {NoStop}%
\bibitem [{\citenamefont {Douglass}\ \emph {et~al.}(2011)\citenamefont {Douglass}, \citenamefont {Land}, \citenamefont {Matthews},\ and\ \citenamefont {Hyde}}]{douglass2011dust}%
  \BibitemOpen
  \bibfield  {author} {\bibinfo {author} {\bibfnamefont {A.}~\bibnamefont {Douglass}}, \bibinfo {author} {\bibfnamefont {V.}~\bibnamefont {Land}}, \bibinfo {author} {\bibfnamefont {L.}~\bibnamefont {Matthews}}, \ and\ \bibinfo {author} {\bibfnamefont {T.}~\bibnamefont {Hyde}},\ }\href@noop {} {\bibfield  {journal} {\bibinfo  {journal} {Physics of Plasmas}\ }\textbf {\bibinfo {volume} {18}},\ \bibinfo {pages} {083706} (\bibinfo {year} {2011})}\BibitemShut {NoStop}%
\bibitem [{\citenamefont {Douglass}\ \emph {et~al.}(2012)\citenamefont {Douglass}, \citenamefont {Land}, \citenamefont {Qiao}, \citenamefont {Matthews},\ and\ \citenamefont {Hyde}}]{douglass2012determination}%
  \BibitemOpen
  \bibfield  {author} {\bibinfo {author} {\bibfnamefont {A.}~\bibnamefont {Douglass}}, \bibinfo {author} {\bibfnamefont {V.}~\bibnamefont {Land}}, \bibinfo {author} {\bibfnamefont {K.}~\bibnamefont {Qiao}}, \bibinfo {author} {\bibfnamefont {L.}~\bibnamefont {Matthews}}, \ and\ \bibinfo {author} {\bibfnamefont {T.}~\bibnamefont {Hyde}},\ }\href@noop {} {\bibfield  {journal} {\bibinfo  {journal} {Physics of Plasmas}\ }\textbf {\bibinfo {volume} {19}},\ \bibinfo {pages} {013707} (\bibinfo {year} {2012})}\BibitemShut {NoStop}%
\bibitem [{\citenamefont {Konopka}\ \emph {et~al.}(1997)\citenamefont {Konopka}, \citenamefont {Ratke},\ and\ \citenamefont {Thomas}}]{konopka1997central}%
  \BibitemOpen
  \bibfield  {author} {\bibinfo {author} {\bibfnamefont {U.}~\bibnamefont {Konopka}}, \bibinfo {author} {\bibfnamefont {L.}~\bibnamefont {Ratke}}, \ and\ \bibinfo {author} {\bibfnamefont {H.}~\bibnamefont {Thomas}},\ }\href@noop {} {\bibfield  {journal} {\bibinfo  {journal} {Physical review letters}\ }\textbf {\bibinfo {volume} {79}},\ \bibinfo {pages} {1269} (\bibinfo {year} {1997})}\BibitemShut {NoStop}%
\bibitem [{\citenamefont {Carstensen}\ \emph {et~al.}(2010)\citenamefont {Carstensen}, \citenamefont {Greiner},\ and\ \citenamefont {Piel}}]{carstensen2010determination}%
  \BibitemOpen
  \bibfield  {author} {\bibinfo {author} {\bibfnamefont {J.}~\bibnamefont {Carstensen}}, \bibinfo {author} {\bibfnamefont {F.}~\bibnamefont {Greiner}}, \ and\ \bibinfo {author} {\bibfnamefont {A.}~\bibnamefont {Piel}},\ }\href@noop {} {\bibfield  {journal} {\bibinfo  {journal} {Physics of Plasmas}\ }\textbf {\bibinfo {volume} {17}} (\bibinfo {year} {2010})}\BibitemShut {NoStop}%
\bibitem [{\citenamefont {Shukla}\ and\ \citenamefont {Eliasson}(2009)}]{shukla2009colloquium}%
  \BibitemOpen
  \bibfield  {author} {\bibinfo {author} {\bibfnamefont {P.~K.}\ \bibnamefont {Shukla}}\ and\ \bibinfo {author} {\bibfnamefont {B.}~\bibnamefont {Eliasson}},\ }\href@noop {} {\bibfield  {journal} {\bibinfo  {journal} {Reviews of Modern Physics}\ }\textbf {\bibinfo {volume} {81}},\ \bibinfo {pages} {25} (\bibinfo {year} {2009})}\BibitemShut {NoStop}%
\bibitem [{\citenamefont {Shukla}\ and\ \citenamefont {Mamun}(2015)}]{shukla2015introduction}%
  \BibitemOpen
  \bibfield  {author} {\bibinfo {author} {\bibfnamefont {P.~K.}\ \bibnamefont {Shukla}}\ and\ \bibinfo {author} {\bibfnamefont {A.}~\bibnamefont {Mamun}},\ }\href@noop {} {\emph {\bibinfo {title} {Introduction to dusty plasma physics}}}\ (\bibinfo  {publisher} {CRC press},\ \bibinfo {year} {2015})\BibitemShut {NoStop}%
\bibitem [{\citenamefont {Ignatov}(2005)}]{ignatov2005basics}%
  \BibitemOpen
  \bibfield  {author} {\bibinfo {author} {\bibfnamefont {A.}~\bibnamefont {Ignatov}},\ }\href@noop {} {\bibfield  {journal} {\bibinfo  {journal} {Plasma physics reports}\ }\textbf {\bibinfo {volume} {31}} (\bibinfo {year} {2005})}\BibitemShut {NoStop}%
\bibitem [{\citenamefont {Galli}\ and\ \citenamefont {Kortshagen}(2010)}]{galli2010charging}%
  \BibitemOpen
  \bibfield  {author} {\bibinfo {author} {\bibfnamefont {F.}~\bibnamefont {Galli}}\ and\ \bibinfo {author} {\bibfnamefont {U.~R.}\ \bibnamefont {Kortshagen}},\ }\href@noop {} {\bibfield  {journal} {\bibinfo  {journal} {IEEE Transactions on Plasma Science}\ }\textbf {\bibinfo {volume} {38}},\ \bibinfo {pages} {803} (\bibinfo {year} {2010})}\BibitemShut {NoStop}%
\bibitem [{\citenamefont {Gatti}\ and\ \citenamefont {Kortshagen}(2008)}]{gatti2008analytical}%
  \BibitemOpen
  \bibfield  {author} {\bibinfo {author} {\bibfnamefont {M.}~\bibnamefont {Gatti}}\ and\ \bibinfo {author} {\bibfnamefont {U.}~\bibnamefont {Kortshagen}},\ }\href@noop {} {\bibfield  {journal} {\bibinfo  {journal} {Physical Review E}\ }\textbf {\bibinfo {volume} {78}},\ \bibinfo {pages} {046402} (\bibinfo {year} {2008})}\BibitemShut {NoStop}%
\bibitem [{\citenamefont {Allan}\ \emph {et~al.}(2024)\citenamefont {Allan}, \citenamefont {Caswell}, \citenamefont {Keim}, \citenamefont {van~der Wel},\ and\ \citenamefont {Verweij}}]{allan_2024_12708864}%
  \BibitemOpen
  \bibfield  {author} {\bibinfo {author} {\bibfnamefont {D.~B.}\ \bibnamefont {Allan}}, \bibinfo {author} {\bibfnamefont {T.}~\bibnamefont {Caswell}}, \bibinfo {author} {\bibfnamefont {N.~C.}\ \bibnamefont {Keim}}, \bibinfo {author} {\bibfnamefont {C.~M.}\ \bibnamefont {van~der Wel}}, \ and\ \bibinfo {author} {\bibfnamefont {R.~W.}\ \bibnamefont {Verweij}},\ }\href {\doibase 10.5281/zenodo.12708864} {\enquote {\bibinfo {title} {soft-matter/trackpy: v0.6.4},}\ } (\bibinfo {year} {2024})\BibitemShut {NoStop}%
\bibitem [{\citenamefont {Abadi}\ \emph {et~al.}(2015)\citenamefont {Abadi}, \citenamefont {Agarwal}, \citenamefont {Barham}, \citenamefont {Brevdo}, \citenamefont {Chen}, \citenamefont {Citro}, \citenamefont {Corrado}, \citenamefont {Davis}, \citenamefont {Dean}, \citenamefont {Devin}, \citenamefont {Ghemawat}, \citenamefont {Goodfellow}, \citenamefont {Harp}, \citenamefont {Irving}, \citenamefont {Isard}, \citenamefont {Jia}, \citenamefont {Jozefowicz}, \citenamefont {Kaiser}, \citenamefont {Kudlur}, \citenamefont {Levenberg}, \citenamefont {Man\'{e}}, \citenamefont {Monga}, \citenamefont {Moore}, \citenamefont {Murray}, \citenamefont {Olah}, \citenamefont {Schuster}, \citenamefont {Shlens}, \citenamefont {Steiner}, \citenamefont {Sutskever}, \citenamefont {Talwar}, \citenamefont {Tucker}, \citenamefont {Vanhoucke}, \citenamefont {Vasudevan}, \citenamefont {Vi\'{e}gas}, \citenamefont {Vinyals}, \citenamefont {Warden}, \citenamefont {Wattenberg}, \citenamefont {Wicke}, \citenamefont {Yu},\ and\ \citenamefont
  {Zheng}}]{tensorflow2015-whitepaper}%
  \BibitemOpen
  \bibfield  {author} {\bibinfo {author} {\bibfnamefont {M.}~\bibnamefont {Abadi}}, \bibinfo {author} {\bibfnamefont {A.}~\bibnamefont {Agarwal}}, \bibinfo {author} {\bibfnamefont {P.}~\bibnamefont {Barham}}, \bibinfo {author} {\bibfnamefont {E.}~\bibnamefont {Brevdo}}, \bibinfo {author} {\bibfnamefont {Z.}~\bibnamefont {Chen}}, \bibinfo {author} {\bibfnamefont {C.}~\bibnamefont {Citro}}, \bibinfo {author} {\bibfnamefont {G.~S.}\ \bibnamefont {Corrado}}, \bibinfo {author} {\bibfnamefont {A.}~\bibnamefont {Davis}}, \bibinfo {author} {\bibfnamefont {J.}~\bibnamefont {Dean}}, \bibinfo {author} {\bibfnamefont {M.}~\bibnamefont {Devin}}, \bibinfo {author} {\bibfnamefont {S.}~\bibnamefont {Ghemawat}}, \bibinfo {author} {\bibfnamefont {I.}~\bibnamefont {Goodfellow}}, \bibinfo {author} {\bibfnamefont {A.}~\bibnamefont {Harp}}, \bibinfo {author} {\bibfnamefont {G.}~\bibnamefont {Irving}}, \bibinfo {author} {\bibfnamefont {M.}~\bibnamefont {Isard}}, \bibinfo {author} {\bibfnamefont {Y.}~\bibnamefont {Jia}}, \bibinfo
  {author} {\bibfnamefont {R.}~\bibnamefont {Jozefowicz}}, \bibinfo {author} {\bibfnamefont {L.}~\bibnamefont {Kaiser}}, \bibinfo {author} {\bibfnamefont {M.}~\bibnamefont {Kudlur}}, \bibinfo {author} {\bibfnamefont {J.}~\bibnamefont {Levenberg}}, \bibinfo {author} {\bibfnamefont {D.}~\bibnamefont {Man\'{e}}}, \bibinfo {author} {\bibfnamefont {R.}~\bibnamefont {Monga}}, \bibinfo {author} {\bibfnamefont {S.}~\bibnamefont {Moore}}, \bibinfo {author} {\bibfnamefont {D.}~\bibnamefont {Murray}}, \bibinfo {author} {\bibfnamefont {C.}~\bibnamefont {Olah}}, \bibinfo {author} {\bibfnamefont {M.}~\bibnamefont {Schuster}}, \bibinfo {author} {\bibfnamefont {J.}~\bibnamefont {Shlens}}, \bibinfo {author} {\bibfnamefont {B.}~\bibnamefont {Steiner}}, \bibinfo {author} {\bibfnamefont {I.}~\bibnamefont {Sutskever}}, \bibinfo {author} {\bibfnamefont {K.}~\bibnamefont {Talwar}}, \bibinfo {author} {\bibfnamefont {P.}~\bibnamefont {Tucker}}, \bibinfo {author} {\bibfnamefont {V.}~\bibnamefont {Vanhoucke}}, \bibinfo {author}
  {\bibfnamefont {V.}~\bibnamefont {Vasudevan}}, \bibinfo {author} {\bibfnamefont {F.}~\bibnamefont {Vi\'{e}gas}}, \bibinfo {author} {\bibfnamefont {O.}~\bibnamefont {Vinyals}}, \bibinfo {author} {\bibfnamefont {P.}~\bibnamefont {Warden}}, \bibinfo {author} {\bibfnamefont {M.}~\bibnamefont {Wattenberg}}, \bibinfo {author} {\bibfnamefont {M.}~\bibnamefont {Wicke}}, \bibinfo {author} {\bibfnamefont {Y.}~\bibnamefont {Yu}}, \ and\ \bibinfo {author} {\bibfnamefont {X.}~\bibnamefont {Zheng}},\ }\href {https://www.tensorflow.org/} {\enquote {\bibinfo {title} {{TensorFlow}: Large-scale machine learning on heterogeneous systems},}\ } (\bibinfo {year} {2015}),\ \bibinfo {note} {software available from tensorflow.org}\BibitemShut {NoStop}%
\bibitem [{\citenamefont {Gogia}\ \emph {et~al.}(2020)\citenamefont {Gogia}, \citenamefont {Yu},\ and\ \citenamefont {Burton}}]{gogia2020intermittent}%
  \BibitemOpen
  \bibfield  {author} {\bibinfo {author} {\bibfnamefont {G.}~\bibnamefont {Gogia}}, \bibinfo {author} {\bibfnamefont {W.}~\bibnamefont {Yu}}, \ and\ \bibinfo {author} {\bibfnamefont {J.~C.}\ \bibnamefont {Burton}},\ }\href@noop {} {\bibfield  {journal} {\bibinfo  {journal} {Physical Review Research}\ }\textbf {\bibinfo {volume} {2}},\ \bibinfo {pages} {023250} (\bibinfo {year} {2020})}\BibitemShut {NoStop}%
\end{thebibliography}%




\clearpage



\makeatletter 
\onecolumngrid
\renewcommand{\thefigure}{S\@arabic\c@figure}
\setcounter{figure}{0}
\makeatother


\renewcommand{\theequation}{S\arabic{equation}}






\section*{Supplementary Materials: ``Physics-tailored machine learning reveals unexpected physics in dusty plasmas"}


\author
{Wentao Yu, Eslam Abdelaleem, Ilya Nemenman, Justin C. Burton$^{\ast}$\\
\\
\normalsize{Department of Physics, Emory University,}\\
\normalsize{400 Dowman Dr., Atlanta, GA 30322, USA}\\
\\
\normalsize{$^\ast$To whom correspondence should be addressed; E-mail:  justin.c.burton@emory.edu.}
}


\date{}





\subsection*{Details of the model structure}
Our model is implemented in TensorFlow \cite{tensorflow2015-whitepaper}. In this section, all the bold text are functions in TensorFlow, with input parameters in the bracket after the function, if necessary. If parameter values are not mentioned, they are assumed to be the default. As described in the main text, the model consists of 3 neural networks (NNs) trained in parallel: $g_\text{int}, \vec{g}_\text{env}$, and $g_\gamma$. 
Both $g_\text{int}$ and $\vec{g}_\text{env}$ have 3 dense-connected hidden layers, with \textbf{he\_normal} initialization and \textbf{L2} regularization. The network $g_\text{int}$ has 32 neurons for each hidden layer with \textbf{leakyrelu} (alpha = 0.1), \textbf{tanh}, and \textbf{leakyrelu} (alpha = 0.1) as activation functions, respectively. The last hidden layer is fully connected to a single output, the magnitude of the reduced interaction force in the $xy$ plane, multiplied by horizontal separation, $f_{ij}\rho_{ij}$. The multiplication of the force by $\rho_{ij}$ serves two purposes. The first is to lessen the divergence of the output as $\rho_{ij}\to 0$. The second is to save considerable computing time by not calculating a square root for every interaction force vector, which is calculated for each particle interaction pair:
\begin{equation}
    \vec{f}_{ij} = f_{ij}\hat{\rho}_{ij} = \frac{f_{ij}\rho_{ij}\vec{\rho}_{ij}}{\rho_{ij}^2}.
    \label{f_int}
\end{equation}
The network $\vec{g}_\text{env}$ has 16 neurons for each hidden layer with \textbf{elu}, \textbf{tanh}, and \textbf{elu} as activation functions, respectively. The last hidden layer is fully connected to two outputs, $f_{i,x}^\text{env}$ and $f_{i,y}^\text{env}$. Finally, the network $g_\gamma$ has 2 hidden layers with 16 neurons each, and \textbf{elu} and \textbf{tanh} as activation functions, respectively. The last hidden layer is fully connected to a single output: the damping coefficient, $\gamma_i$. As described in the main text, our model fits the reduced net force, $\sum_{j}\vec{f}_{ij}+\vec{f}_i^\text{env} - \gamma_i\dot{\vec{\rho}}_i$, to each particle's experimental acceleration, $\Ddot{\vec{\rho}}_i$. 

To reduce the amplification of measurement error by temporal differentiation, we apply the weak form \cite{gurevich2019robust} in our loss function: 
\begin{equation}
    L = \dfrac{1}{2N_pT_\text{train}}\sum_{i = 0}^{(N_p-1)}\sum_{t\in\mathbb{T}_\text{train}}\sum_\alpha^{\{x,y\}} L_{i,t,\alpha},
    \label{loss}
\end{equation}
\begin{equation}
    L_{i,t,\alpha} = H\left(w\circledast_t(\vec{f}_i^\text{env} + \sum_{j}\vec{f}_{ij} - \gamma_i\dot{\vec{\rho}}_i - \Ddot{\vec{\rho}}_i)_\alpha;\delta\right)
    \label{lossLK}
\end{equation}
Here $T_\text{train}$ is the total number of frames for the particle trajectories in the training dataset, $\mathbb{T}_\text{train}$, and $w$ is a customized weight function, defined in the range $[-\tau\Delta/2, \tau\Delta/2]$:
\begin{equation}
    w_{t'} = w(t'\Delta ) = \frac{30}{(\tau\Delta)^5}\left((t'\Delta )^2 - (\tau\Delta/2)^2\right)^2,
\end{equation}
where the recording time step $\Delta = 0.005$ s, and $\tau = 16$ is the size of the convolution window. The function $H$ is a Huber loss function that reduces the relative weight of outliers in the loss function. The parameter $\delta$ controls the threshold of this reduction. The convolution function $\circledast_t$ is defined as: 
\begin{equation}
    a\circledast_tb = \Delta \int_{-\tau/2}^{\tau/2} a(t'\Delta)b(t'\Delta+t\Delta)\,dt'= \Delta \sum_{t'=-\tau}^{\tau}S_{t'}a_{t'}b_{t+t'}.
\label{conv}
\end{equation}
In the last step of the equation above, Simpson discretization is used to compute the integral over each window, with the coefficient: 
\begin{equation}
    S_{t'}= 
\begin{cases}
    1/3, & \text{if } |t'|=\tau\\
    4/3, & \text{if } |t'|<\tau\text{ and }(t'+\tau) \text { is odd}\\
    2/3, & \text{if } |t'|<\tau\text{ and }(t'+\tau) \text{ is even}\\
    0, & \text{else}.
\end{cases}
\end{equation}
By definition, at $t' = \pm \tau/2$, $w(t') = 0$ and $\dot w(t') = 0$. Therefore, it is easily proven through integration by parts that: 
\begin{equation}
    w\circledast_t\dot{\vec{\rho}}_i = -\dot w\circledast_t\vec{\rho}_i
\end{equation}
\begin{equation}
    w\circledast_t\Ddot{\vec{\rho}}_i = \Ddot w\circledast_t\vec{\rho}_i
\end{equation}
As a result, our loss function becomes: 
\begin{equation}
     L_{i,t,\alpha} = H\left((w\circledast_t\vec{f}_i^\text{env} + \sum_{j}w\circledast_t\vec{f}_{ij} + \gamma_i \Dot w\circledast_t{\vec{\rho}}_i -  \Ddot w\circledast_t{\vec{\rho}}_i)_\alpha;\delta\right)
\end{equation}
Thus, by using the weak form, temporal derivatives of experimental particle positions are replaced by derivatives of the weight function, which is analytic. 

As mentioned previously, the parameter $\delta$ controls the crossover from quadratic to linear loss in the Huber loss function. When $x < \delta$, $H(x; \delta) \propto x^2$ and when $x > \delta$, $H(x; \delta)\propto x$. Considering that a very large fitting error on a single data point might arise from other sources of noise (for example, tracking error), this large error should be deemphasized (only matter linearly) in our loss function. The parameter $\delta$ is chosen to be:
\begin{equation}
    \delta = 0.25\sqrt{\text{TSS}_D} = 0.25\sqrt{\dfrac{1}{2N_pT_D}\sum_{i = 0}^{(N_p-1)}\sum_{t\in\mathbb{T}_D}\sum_\alpha^{\{x,y\}} (\Ddot w\circledast_t\vec{\rho}_i)^2_\alpha},
    \label{deltaguess}
\end{equation}
where TSS is total sum of squares of the experimental acceleration in the loss function. $D$ refers to either train or test data set. To quantify the quality of the model's fit, we define $R^2$ as: 
\begin{equation}
    R^2 = 1 - \frac{\text{RSS}_\text{test}}{\text{TSS}_\text{test}},
\end{equation}
where RSS is residual sum of squares:

\begin{equation}
    \text{RSS}_D = \dfrac{1}{2N_pT_D}\sum_{i = 0}^{(N_p-1)}\sum_{t \in \mathbb{T}_D}\sum_\alpha^{\{x,y\}}(w\circledast_t\vec{f}_i^\text{env} + \sum_{j}w\circledast_t\vec{f}_{ij} + \gamma_i \Dot w\circledast_t{\vec{\rho}}_i -  \Ddot w\circledast_t{\vec{\rho}}_i)_\alpha^2.
\end{equation}
We note that for $R^2 > 0.99$, the average percentage error should be $\sqrt{1 - R^2} < 10\%$. Therefore, we set an arbitrary threshold, $\delta$ = 0.25, which indicates that data with an error that is 2.5 times the average error should be considered an outlier in the Huber loss. Finally, the data is split into 10 temporal sections, and 10 models are trained by 10-fold cross-validation. Such splitting ensures that models inferred in one section work in the others, so that there are no significant drifts in the experiments. Note that because of the convolution (Eq.\ \ref{conv}), for a data with time length $T$, $t$ can only be defined on $\tau/2 \leq t < T-\tau/2$. For the $l$-th model, 
$\mathbb{T}_\text{test} =\left\{t|\tau/2 + \frac{l-1}{10}(T-\tau)\leq t<\tau/2 + \frac{l}{10}(T-\tau)\right\}$, and 
$\mathbb{T}_\text{train} = \left\{
t|\tau/2\leq t<T-\tau/2
\text{ and } t\notin\mathbb{T}_\text{test}\right\}$.
The average (test) $R^2$ of the 10 models for the 10-fold validation is reported in Table~1. The error bars shown in Figs.~3, 4 in the {\em Main text} are calculated from the standard deviation of the 10 models' prediction. We note that this estimation of the error bars only includes the variance of the model, plus the variation caused by the temporal plasma environment fluctuation in the experiments, while the bias of the model is excluded. 

{\color{black}The time it takes to fully train our model scales as $N_p^2$ and typically takes 2-3 hours on desktop computer with an Intel 14900 processor. At the cost of accuracy and more complex book-keeping (see {\bf Data processing} below), one can simply truncate the interaction force between two particles at large separations, so that the training time scales as $N_p$. However, this is not the most challenging part when extending our model to a large number of particles. The model's ability to infer forces requires a particle-level identifier, $s_i$, meaning mis-identification or mis-tracking of particles can have a detrimental effect on the model's performance.}  

\subsection*{Data processing}

To train the model, the data $x_{i,t}$, $y_{i,t}$, $z_{i,t}$, and $s_{i,t}$ needs to be organized into a form that can be efficiently iterated over to save computational time. Note that $s_i$ is a time-averaged identifier of particle $i$ and is independent of $t$. However, for consistency in the input, we constructed the array of $s_{i,t} \equiv s_i$ at all times. We need three tensors that can be used to calculated the convolution of the data with $w$, $\dot w$, and $\ddot w$. Thus, each term in the loss function was associated with a separate tensor of data to compute the convolution. The data is first processed into three tensors $X^0$, $X^1$, and $Y$. $Y$ is the target, which is a 3D tensor with shape of $N_p\times (T-\tau)\times 2$. $Y_{i,t,\alpha} = \Ddot w\circledast_t\alpha_i$ where $\alpha$ is either $x$ or $y$. Similarly, $X^1_{i,t,\alpha} = \dot w\circledast_t\alpha_i$. $X^0$ is a 5D tensor, with a shape of $N_p\times (T-\tau)\times(\tau-1)\times N_p\times 4$ . To explain the meaning of $X^0_{i,t,t',k,\alpha}$, we first define an index function on $0 \leq i< N_p$, $0\leq k < N_p$:
\begin{equation}
    n(i,k)= 
\begin{cases}
    i, & \text{if } k = 0,\\
    k-1, & \text{if } 0 < k \leq i,\\
    k, & \text{if } i < k < N_p.
\end{cases}
\end{equation}
Then $X^0_{i,t,t',k,\alpha} = \alpha_{j,t+t'+1}$, where $\alpha$ can be $x$, $y$, $z$, or $s$, and $j = n(i,k)$. Here, note that when calculating $w\circledast_t f_{ij}$, only the input from time $t-\tau/2+1$ to $t+\tau/2-1$ is needed, with a total length of $\tau-1$, because $w_{\pm\tau/2} = 0$.  Finally, the first two dimensions of all three tensors are flattened, and the last two dimensions of $X^0$ are flattened, making $X^0, X^1$ and $Y$ 3D, 2D, and 2D tensors, respectively.
    \label{fiteq}

\subsection*{Dusty plasma simulations}

In order to test the accuracy of the ML methods, and the inference of the mass and charge of particles, we simulated our dusty plasma system using a custom molecular dynamics code. The simulations are similar to those used in previous studies \cite{gogia2017emergent,gogia2020intermittent,yu2022extracting}. The simulations consisted of 15 spherical particles whose diameters were chosen from a Gaussian distribution with a mean of $d_0$ = 10 $\mu$m and a standard deviation of 1 $\mu$m. In the horizontal, $xy$-plane, the particles were confined by a harmonic potential with a small degree of asymmetry to match the experiments. They also experienced a vortical force to induce rotation of the system, leading to the following environmental reduced force:
\begin{align}
f^{\text{env}}_{i,x}&=(1+\beta) \chi_h q_i x_i/m_i+\Omega^2 y_i-\gamma \dot{x}_i, \\
f^{\text{env}}_{i,y}&=(1-\beta) \chi_h q_i y_i/m_i-\Omega^2 x_i-\gamma \dot{y}_i. 
\end{align}
The degree of asymmetry of the potential was determined by the dimensionless number $\beta$, $\chi_h$ is the electric field gradient, $q_i$ and $m_i$ are the charge and mass of particle $i$, $\Omega$ is the strength of the background vorticity from ion drag, and $x_i$ and $y_i$ are the horizontal coordinates of particle $i$. {\color{black}These parameters are all necessary to describe the general environmental confinement of dust particles, as discussed in detail in Ref.~\cite{yu2022extracting}}. Dotted variables indicate differentiation with respect to time and the Epstein drag force is determined by $\gamma$. The mass of each particle was computed as $m_i=\rho_p\pi d_i^3/6$, where $\rho_p$ = 1,510 kg$\cdot$m$^{-3}$, and $d_i$ is the diameter of particle $i$.

In the vertical direction, the particles experienced forces due to a linearly-varying electric field, and gravity. The reduced force
was determined by the following equation:
\begin{align}
f^{\text{env}}_{i,z}&=\text{min}(E_0+\chi_z z_i,0) q_i/m_i-g-\gamma \dot{z}_i+\eta w(t).
\end{align}
Here $E_0$ is a constant vertical electric field, $\chi_z$ is the electric field gradient, $z_i$ is the vertical position of the particle, and $g$ = 9.81 m$\cdot$s$^{-2}$ is the acceleration due to gravity. The \textbf{min} function guarantees that the electric force will never change sign, and thus the edge of the plasma sheath occurs at $z_\text{edge}=-E_0/\chi_z$, a small distance above $z=0$. The last term provides a small amount of stochastic noise in the $z$ direction. This noise drives oscillations in $z$ since the particles behave as stochastic harmonic oscillators with a well-defined resonance frequency. The function $w(t)$ represents a Wiener process with zero mean and unit standard deviation, and $\eta$ is the strength of the noise. Since we are not inferring forces in $z$, this does not affect the inference procedure, and is based on previous experiments in our lab illustrating $z$ oscillations originating from Brownian motion \cite{yu2022extracting} and spontaneous oscillations due to delayed charging at low pressures \cite{harper2020origin}. We also allow the charge on the particle to vary linearly within the sheath, increasing in magnitude as $z_i$ decreases. This was done by treating each particle as a spherical capacitor, and parameterizing the charge in the following way:
\begin{align}
q_i=\text{min}(2\pi\epsilon_0 d_i V (1-z_i/l_q), -8\times10^{-16}),
\label{simq}
\end{align}
where the units of charge are in Coulombs. This guarantees that the magnitude of the (negative) charge on the particle will never be smaller that 5,000e, and the magnitude of charge increases deeper into the sheath (smaller $z_i$). Here $l_q$ is a length scale that determines the strength of charge variation in the sheath. The voltage $V$ is a constant that determines the charge on a particle at $z=0$.

The parameters described here, such as electric field, are difficult to relate to experimental measurements. Thus, we fixed these parameters by relating them to the typical frequencies of small oscillations of the particles around their equilibrium positions. Experimentally, these can be measured from the 3D tracking data \cite{yu20233d}, and are given by $\omega_h$ in the horizontal direction, and $\omega_z$ in the vertical direction. Linearizing the force around $z=0$, so that $f^{\text{env}}_{z}=0$, $\omega_h^2=-df^{\text{env}}_{x}/dx$, and $\omega_z^2=-df^{\text{env}}_{z}/dz$, we arrive at the following relationships:
\begin{align}
\chi_h&=-\dfrac{\rho_p d_0^2\omega_h^2}{12V\epsilon_0}, \\
E_0&=\dfrac{\rho_p d_0^2g}{12V\epsilon_0},\\
\chi_z&=\dfrac{\rho_p d_0^2(g-l_q\omega_v^2)}{12l_qV\epsilon_0}.
\end{align}
This way a particle with diameter $d_0$ would have its equilibrium position at $z=0$, and frequencies of small oscillations exactly equal to $\omega_h$ and $\omega_z$. However, since particle sizes are drawn from a Gaussian distribution centered at $d_0$, the frequencies vary as well. 

In addition to the environmental forces, the particles experienced a pairwise, non-reciprocal repulsive force. This force stems from basic Coulomb repulsion, but also from the wake of ions streaming past each particle. As done in Ref.\ \cite{kryuchkov2020strange}, we parameterized this ion wake by an effective positive cloud of charge with magnitude $\Tilde{q}q_i$ at a distance $h$ beneath each particle. The force between particles was derived from the following potential:
\begin{align}
\phi(\vec{r})=\dfrac{q_iq_j}{4\pi\epsilon_0\lambda_D}\left[\dfrac{e^{-r/\lambda_D}}{r/\lambda_D}-\Tilde{q}\dfrac{e^{-r_w/\lambda_D}}{r_w/\lambda_D}\left(1+b\dfrac{e^{-r_w/\lambda_D}}{r_w/\lambda_D}\right)^{-1}\right].
\label{nonrecippot}
\end{align}
Here, $\phi(\vec{r})$ is the potential of the $i$th particle in the field of the $j$th particle and its wake, and $f_{ij}=-\vec{\nabla}_i\phi$. The position vector between the particles is $\vec{r}$, $r_w=|\vec{r}-h\hat{z}|$ is the distance from particle $i$ to the wake of particle $j$, $\hat{z}$ is the unit vector in the $z$ direction, $\lambda_D$ is the Debye screening length in the plasma, and $b$ is dimensionless cutoff used to truncate the divergence of the wake interaction since the wake is not a point charge, but more of a cloud. With these environmental and interaction forces, the Newton's 2\textsuperscript{nd} law was integrated forward in time using the 2\textsuperscript{nd}-order velocity Verlet method.

Without energy input, Epstein drag would drain the energy from the system and the particles would assume equilibrium positions. However, there are three mechanisms that drive kinetic and potential energy into the particles' motion. The first is the vortical force from ion drag, which is non-conservative. The second is the small amount of stochastic noise in the $z$-direction. The third is the non-reciprocal interaction force (also non-conservative) \cite{ivlev2015statistical}. {\color{black} For a given simulation, we chose parameters that produced particle motion most visually similar to the experiments, or parameters that could be measured directly from experiments (like oscillation frequency).} Movie S6 shows that the resulting motion of the particles indeed looks strikingly similar to the experiments, and can be easily analyzed by our ML model. 

Prior to training the model, Gaussian-distributed measurement error with standard deviation 0.005 mm was added to each particle position to simulate experimental particle tracking error. In our simulation, we used $\lambda_D$ = 0.8 mm (Eq.~\ref{nonrecippot}) for all particles. Figure~\hyperref[sp2]{S2\textit{A}} shows that at the same $z$-position, there is only a weak dependence of the fitted effective screening length on different particle sizes since $\lambda$ only varies from 0.48 - 0.52 mm for different particle pairs. This indicates that the particle-dependent effective screening length $\lambda$ in experiments is real (Fig.~4\textit{A}), rather than an artifact of the ML model. Moreover, the predicted interaction agrees with the exact interaction with less than 10\% error (Fig.~\hyperref[sp2]{S2\textit{A}}). Even though the fit is very good, as discussed in the main text, the presence of a virtual ion wake can systematically reduce the fitted values of the screening length ($\lambda = 0.52$ mm from the fit, and  $\lambda_D = 0.80$ mm in Eq.~\ref{nonrecippot}). 

{\color{black}Figure \hyperref[sp2]{S2\textit{B}} shows the inferred masses from the damping term, assuming Epstein drag (Eq.~2 in the {\em Main Text}), and the mass inferred from the fitting procedure, versus the actual masses of particles used in the simulation. The agreement is remarkable and demonstrates that our model can accurately infer each term in the equation of motion. Figure \hyperref[sp2]{S2\textit{C}} shows the inferred charge on each particle versus the inferred mass. The fitted slope of $p$ = 0.31 is close to the expected value from the simulation, $p$ = 1/3, and reflects the fact that particles at the same vertical position will have the same floating potential, independent of their mass (Eq.~\ref{simq}). However, fitting to Eq.~4 in the {\em Main Text} results in a deviation of the prefactor $A$ = 89 mm$^3\cdot$s$^{-2}$ from the actual $q_1q_2/4\pi\epsilon_0m_1$ = 103 mm$^3\cdot$s$^{-2}$. This deviation of $A$ can cause the inferred $q$ to be systematically lower than the actual $q$ by 5-10\%.

Taken together, Fig.~\hyperref[sp2]{S2} suggests that the inference of interaction forces in simulated data is excellent, ion wake-mediated interactions can significantly reduce the effective screening length, and the inference of particle charge is very good (5-10\% error). 
}


\subsection*{Supplementary movies}
Movies S1-S5 show the 3D motion of the particles in our 5 experiments, as labeled in Table 1 in the main text. Movies S6 and S7 show rotating crystal states with 6 and 9 particles, respectively, which were often observed when the particle number $N\lesssim9$. Movie S8 shows the 3D motion of our dusty plasma simulation. 

\subsection*{Data availability}
All 5 experimental 3D trajectories and simulated trajectories, plus the code of our machine-learning algorithm, is available on github: https://github.com/wyu54/many-body-force-infer

\end{document}